\documentclass[11pt,twoside]{article}
\usepackage{jeffe,fullpage,epsfig,myconcrete,euler,url}
\usepackage{psfrag,subfigure}
\newtheorem{lemma}{Lemma}[section]
\newtheorem{theorem}[lemma]{Theorem}
\newtheorem{corollary}[lemma]{Corollary}
\numberwithin{figure}{section}
\numberwithin{table}{section}

\urldef{\OA}\url{oaich@igi.tu-graz.ac.at}
\urldef{\DB}\url{bremner@unb.ca}
\urldef{\CC}\url{ccortes@cica.es}
\urldef{\VD}\url{vida@cs.mcgill.ca}
\urldef{\ED}\url{eddemaine@uwaterloo.ca}
\urldef{\JE}\url{jeffe@cs.uiuc.edu}
\urldef{\FH}\url{hurtado@ma2.upc.es}
\urldef{\HM}\url{henk@cs.queensu.ca}
\urldef{\MO}\url{markov@cs.uu.nl}
\urldef{\BP}\url{b.palop@escet.urjc.es}
\urldef{\SR}\url{rsuneeta@crab.rutgers.edu}
\urldef{\VS}\url{vera@ma2.upc.es}
\urldef{\MS}\url{soss@cs.mcgill.ca}
\urldef{\GT}\url{godfried@cs.mcgill.ca}
\urldef{\paperURL}\url{http://www.uiuc.edu/~jeffe/pubs/flipturn.html}

\begin{document}

\pagestyle{myheadings}

\markboth{Flipturning Polygons}{Bellairs Polygonal Entanglement Workshop Group}

\begin{titlepage}

\title{\unskip\textbf{Flipturning Polygons}%
	\thanks{Portions of this work were done at the Workshop on
	Polygonal Entanglement Theory held February 4--11, 2000 at
	McGill University's Bellairs Research Center, Holetown,
	Barbados.  For the most recent version of this paper, see
	\paperURL.}}

\def\oneauthor#1#2#3{\small\begin{tabular}{c}\large#1\\#2\\#3\end{tabular}}

\author{\begin{tabular}{c@{}cc@{\quad}}
	\oneauthor{Oswin Aichholzer\thanks{Supported by the
		Austrian Programme for Advanced Research and
		Technology (APART).}} 
		{Technische \Universitat\ Graz}{\OA}
	&
	\oneauthor{Carmen \Cortes}{Universidad de Sevilla}{\CC}
	&
	\oneauthor{Erik D. Demaine}{University of Waterloo}{\ED}
	\\[5ex]
	\oneauthor{Vida \Dujmovic}{McGill University}{\VD}
	&
	\oneauthor{Jeff Erickson\thanks{Contact author.
		Partially supported by a Sloan Fellowship.}} 
		{University of Illinois}{\JE}
	&
	\oneauthor{Henk Meijer}{Queen's University}{\HM}
	\\[5ex]
	\oneauthor{Mark Overmars}{Universiteit Utrecht}{\MO}
	&
	\oneauthor{\Belen\ Palop\thanks{Partially supported by Proyecto
	DGES-MEC-PB98-0933.}}{Universidad Rey Juan Carlos}{\BP}
	&
	\oneauthor{Suneeta Ramaswami}{Rutgers University}{\SR}
	\\[5ex]
	&
	\oneauthor{Godfried T. Toussaint\thanks{Partially
	supported by NSERC Grant No.\ A9293.}}{McGill University}{\GT}
	\\[5ex]
\end{tabular}
}

\maketitle
\begin{abstract}\noindent
A \emph{flipturn} is an operation that transforms a nonconvex simple
polygon into another simple polygon, by rotating a concavity 180
degrees around the midpoint of its bounding convex hull edge.  Joss
and Shannon proved in 1973 that a sequence of flipturns eventually
transforms any simple polygon into a convex polygon.  This paper
describes several new results about such flipturn sequences.  We show
that any orthogonal polygon is convexified after at most $n-5$
arbitrary flipturns, or at most $\floor{5(n-4)/6}$ well-chosen
flipturns, improving the previously best upper bound of $(n-1)!/2$.
We also show that any simple polygon can be convexified by at most
$n^2-4n+1$ flipturns, generalizing earlier results of Ahn \etal\@
These bounds depend critically on how degenerate cases are handled; we
carefully explore several possibilities.  We describe how to maintain
both a simple polygon and its convex hull in $O(\log^4 n)$ time per
flipturn, using a data structure of size $O(n)$.  We show that
although flipturn sequences for the same polygon can have
significantly different lengths, the shape and position of the final
convex polygon is the same for all sequences and can be computed in
$O(n\log n)$ time.  Finally, we demonstrate that finding the longest
convexifying flipturn sequence of a simple polygon is NP-hard.
\end{abstract}

\thispagestyle{empty}
\setcounter{page}{0}
\end{titlepage}

\section{Introduction}

A central problem in polymer physics and molecular biology is the
reconfiguration of large molecules (modeled as polygons) such as
circular DNA~\cite{kamenetskii-97}.  Most of the research in this area
involves computer-intensive Monte-Carlo simulations.  To simplify
these simulations they are usually restricted to the integer lattices
$\Z^2$ and $\Z^3$, although some work has also been done on the FCC
lattice~\cite{rensburg-90}.  Like the related algorithmic robotics
research on linkages, the problems of interest to physicists and
biologists involve closed simple polygons~\cite{dubins-88}, open
simple polygonal chains~\cite{mcmillan-79} and simple polygonal
trees~\cite{finn-96}, \ie, polygons, chains, and trees that do not
intersect themselves; hence the term \emph{self-avoiding walks} for
the case of polygons and chains.  Generating a random self-avoiding
walk from scratch is difficult, especially if it must return to its
starting point as in the case of polygons.  The waiting time is too
long due to attrition; if a random walk crosses itself at any point
other than its starting point, it must be discarded and a new walk
started.  Therefore an efficient method frequently used to generate
random chains or polygons is to modify one such object into another
using a simple operation called a \emph{pivot}.  Unlike the work in
linkages, however, here we do not care if intersections happen
\emph{during} the pivot as long as when the pivot is complete we end
up with a simple polygon or chain.  In other words, pivots are seen as
instantaneous combinatorial changes, not continuous processes.  In
general the pivots used are selected from a large variety of
transformations such as reflections, rotations, or `cut and paste'
operations on certain subchains.  We refer the reader to a multitude
of such problems and results in~\cite{madras-93}.  For example, Madras
and Sokal~\cite{madras-88} have shown that for all $d\ge 2$, every
simple lattice polygonal chain with $n$ edges in $\Z^d$ can be
straightened by some sequence of at most $2n-1$ suitable pivots while
maintaining simplicity after each pivot.  The pivots used here are
either reflections through coordinate hyperplanes or rotations by
right angles.

In order to prove the ergodicity of their self-avoiding walks, polymer
physicists are interested in convexifying polygons (and straightening
open polygonal chains).  If a polygon can be transformed to some
canonical convex configuration, then any simple polygon can be
reconfigured to any other via this intermediate position.  This
theoretical aspect of polymer physics research resembles the
algorithmic robotics work on convexification of polygonal linkages.
We refer the reader to survey papers of O'Rourke~\cite{o-fucg-00} and
Toussaint~\cite{toussaint-castellon-99} for further references in the
latter area.

\begin{figure}[htb]
\centerline{\epsfig{file=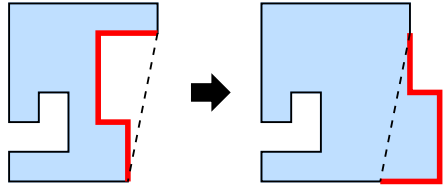,height=1in}}
\caption{A flipturn.  The edges of the pocket are bold (red), and its
lid is dashed.}
\label{Fig/1flipturn}
\end{figure}

\begin{figure}[htb]
\centerline{\epsfig{file=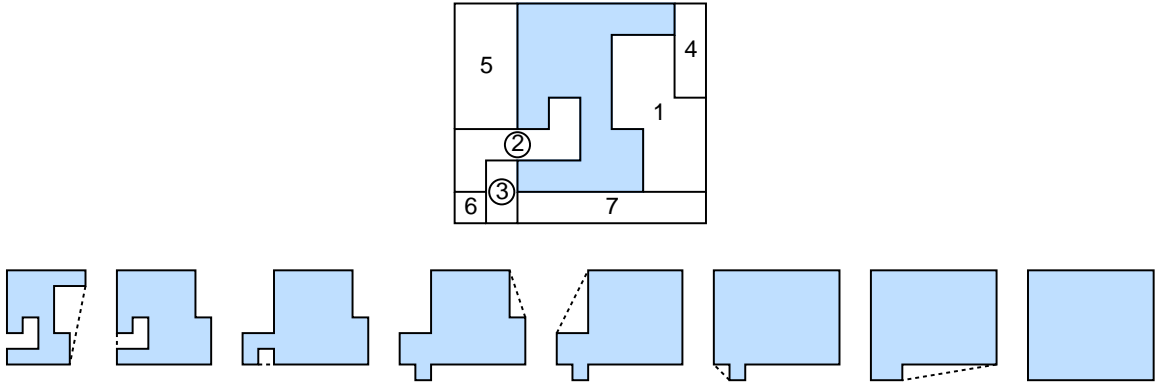,height=2in}}
\caption{A convexifying flipturn sequence.}
\label{Fig/sequence}
\end{figure}

In this paper, we are concerned with one type of pivot of central
concern in polymer physics research.  This pivot is usually called an
\emph{inversion} in the physics literature, but since it seems to have
been first proposed in an unpublished 1973 paper of Joss and
Shannon~\cite{grunbaum-95}, we will follow their terminology and call
it a \emph{flipturn}.  Flipturns are defined as follows.  Any
nonconvex polygon has at least one concavity, or \emph{pocket}.
Formally, a pocket of a nonconvex polygon $P$ is a maximal connected
sequence of polygon edges disjoint from the convex hull of $P$ except
at its endpoints.  The line segment joining the endpoints of a pocket
is called the \emph{lid} of the pocket.  A flipturn rotates a pocket
180 degrees about the midpoint of its lid, or equivalently, reverses
the order of the edges of a pocket without changing their lengths or
orientations.  Figure~\ref{Fig/1flipturn} shows the effect of a single
flipturn on a nonconvex orthogonal polygon, and
Figure~\ref{Fig/sequence} shows a sequence of flipturns transforming
this polygon into a rectangle.  We will illustrate such sequences by
overlaying the resulting polygons and labeling the area added by each
flipturn by its position in the sequence.  (The circled numbers will
be explained in Section~\ref{degen}.)

\subsection{Previous and Related Results}

Joss and Shannon proved that any simple polygon with $n$ sides can be
convexified by a sequence of at most $(n-1)!$ flipturns, by observing
that each flipturn produces a new cyclic permutation of the edges.
Since each flipturn increases the polygon's area, each of the $(n-1)!$
cyclic permutations can occur at most once.  We can immediately
improve this bound to $(n-1)!/2$ by observing that at most half of the
$(n-1)!$ cyclic permutations describe a simple polygon with the proper
orientation.  Although this is the best bound known, it is extremely
loose; Joss and Shannon conjectured that $n^2/4$ flipturns are always
sufficient.  \Grunbaum\ and Zaks~\cite{grunbaum-98} showed that even
crossing polygons could be convexified with a finite number of
flipturns.  Biedl \cite{biedl-2000} discovered a family of polygons
that are convexified only after $(n-2)^2/4$ badly chosen flipturns,
nearly matching Joss and Shannon's conjectured upper bound.  Ahn
\etal~\cite{abchkm-fyl-00} recently proved that any simple polygon can
be convexified by a sequence of at most $n(n-3)/2$ so-called
\emph{modified} flipturns (which we define in Section~\ref{degen}).
Better results are known for orthogonal and lattice polygons in the
plane.  Dubins \etal\ \cite{dubins-88} showed that any simple lattice
polygon in the plane can be convexified with $n-4$ well-chosen
flipturns~\cite{madras-93}.  Until very recently this was the best
upper bound known.  Ahn \etal~\cite{abchkm-fyl-00} show that any
polygon with $s$ distinct edge slopes can be convexified by
$\ceil{n(s-1)/2 - s}$ modified flipturns\footnote{Ahn
\etal~\cite{abchkm-fyl-00} omit the ceiling, so their stated bound is
off by one when $n$ is odd and $s$ is even.}; in particular, $n/2 - 2$
modified flipturns suffice to convexify any orthogonal polygon.

\begin{figure}[htb]
\centering\footnotesize\sf
\begin{tabular}{c}\epsfig{file=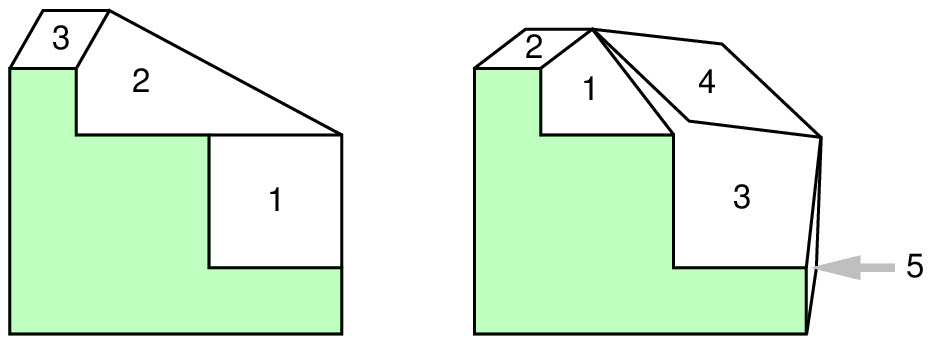,height=1in}\\(a)\end{tabular}
\begin{tabular}{c}\epsfig{file=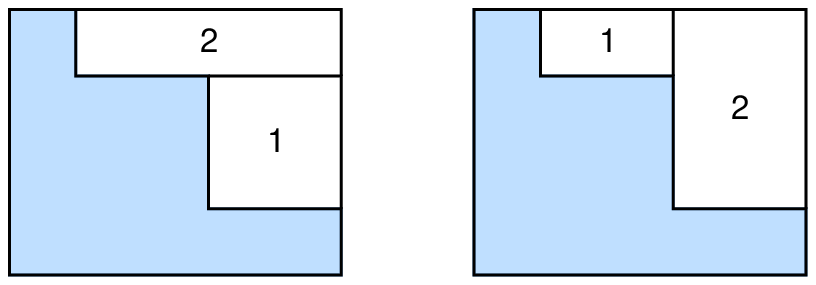,height=1in}\\(b)\end{tabular}
\caption{(a) Different \Erdos-Nagy flip sequences can lead to
different convex shapes.  (b) Different flipturn sequences always lead
to the same convex shape.}
\label{Fig/flip}
\end{figure}

There are significant differences between flipturns and another very
common pivoting rule, the \emph{\Erdos-Nagy flip} \cite{erdos-35,
grunbaum-95, nagy-39, toussaint-erdos-99}, in which a pocket is
\emph{reflected} across its lid.  As with flipturns, any convex
polygon can be convexified using a finite number of flips.  Unlike
flipturns, however, the number of flips required is not bounded by any
function of $n$; in particular, Joss and Shannon constructed a family
of quadrilaterals that require an unbounded number of flips to
convexify \cite{grunbaum-95}.  Another important difference is that
flipturns preserve the \emph{orientation} of polygon edges, while
flips preserve their \emph{order} around the polygon.  This implies
that starting from the same simple polygon, different sequences of
flips can lead to different convex polygons---see Figure
\ref{Fig/flip}(a) for an example---but different flipturn sequences
always lead to the same convex shape.  For further results on both
flips and flipturns for general polygons, simpler algorithms, and a
more complete history of the problem, see~\cite{toussaint-erdos-99}.

\subsection{New Results}

Our results depend critically on the behavior of flipturns in
degenerate cases.  In Section \ref{degen}, we offer three alternate
definitions: \emph{standard}, \emph{extended}, and \emph{modified}
flipturns.  As our naming suggests, we believe that standard flipturns
are closest to the original definition of Joss and Shannon; modified
flipturns were introduced by Ahn \etal~\cite{abchkm-fyl-00}.

In Section \ref{ortho}, we develop a number of new results concerning
convexifying flipturn sequences for orthogonal polygons.  We show that
$\floor{5(n-4)/6}$ well-chosen (standard) flipturns are sufficient,
and $\floor{3(n-4)/4}$ flipturns are sometimes necessary, to convexify
any orthogonal polygon.  We also show that any orthogonal polygon is
convexified after at most $n-5$ \emph{arbitrary} flipturns, and that
some polygons can survive $\floor{5(n-4)/6}$ flipturns.  Finally, we
show that the shortest and longest flipturn sequences for the same
orthogonal polygon can differ in length by at least $(n-4)/4$.
Similar results are derived for extended flipturns.  All of these
bounds improve the previously best known results.  Using techniques
developed in Section \ref{ortho}, we prove in Section \ref{general}
that any polygon can be convexified after at most $n^2-4n+2$ standard
or extended flipturns, generalizing the modified flipturn results of
Ahn \etal~\cite{abchkm-fyl-00}.  Our new upper and lower bounds are
summarized in the first two rows of Tables~\ref{results} and
\ref{results2}; the last row of each table gives the corresponding
results of Ahn \etal for modified flipturns.

\begin{table}[htb]
\centering\small
\begin{tabular}{c|c|c}
{Flipturn type}
		& {Shortest flipturn sequence}
		& {Longest flipturn sequence}
\\[0.25ex]\hline&&\\[-2.5ex]
{standard}
	 	& $\floor{3(n-4)/4} \le ?? \le \floor{5(n-4)/6}$
		& $\floor{5(n-4)/6} \le ?? \le n-5$
\\[0.25ex]
{extended}
		& $\floor{3(n-4)/4}$
		& $\floor{3(n-4)/4} \le ?? \le n-5$
\\[0.25ex]
{modified} \cite{abchkm-fyl-00}
		& $(n-4)/2$
		& $(n-4)/2$
\\[0.25ex]
\end{tabular}
\caption{Bounds for shortest and longest flipturn sequences for
orthogonal polygons.  See Section~\ref{ortho}.}
\label{results}
%
%
\vspace{\floatsep}
\begin{tabular}{c|cc}
{Flipturn type}
		& {$s$-oriented polygons}
		& {arbitrary polygons}
\\[0.25ex]\hline\\[-2.5ex]
{standard}
		& $ns - \floor{(n+5s)/2} - 1$
		& $n^2 - 4n + 1$
\\[0.25ex]
{extended}
		& $ns - \floor{(n+5s)/2} - 1$
		& $n^2 - 4n + 1$
\\[0.25ex]
{modified}  \cite{abchkm-fyl-00}
		& $\ceil{n(s-1)/2} - s$
		& $n(n-3)/2$
\\[0.25ex]
\end{tabular}
\caption{Upper bounds for longest flipturn sequences of more general
polygons.  See Section~\ref{general}.}
\label{results2}
\end{table}

Section \ref{algo} describes how to maintain both a simple polygon and
its convex hull in $O(\log^4 n)$ time per flipturn, using a data
structure of size $O(n)$.  Our data structure is a variant of the
dynamic convex hull structure of Overmars and van Leeuwen
\cite{ol-mcp-81}.  Together with the results of the previous sections,
this implies that we can compute a convexifying sequence of flipturns
for any polygon in $O(n^2\log^4 n)$ time, or for any orthogonal
polygon in $O(n\log^4 n)$ time.

In Section \ref{place}, we prove that for any simple polygon, every
sequence of flipturns eventually leads to the same convex polygon, and
we can compute this convex polygon in $O(n\log n)$ time.  As we
already mentioned, the fact that the \emph{shape} of the final convex
polygon is independent of the flipturn sequence is rather obvious, but
the independence of the final polygon's \emph{position} requires
considerably more effort.

Finally, in Section \ref{worst}, we show that finding the longest
flipturn sequence for a given simple polygon is NP-hard.

\section{The Importance of Being Degenerate}
\label{degen}

The behavior of flipturn sequences depends critically on how flipturns
are defined in degenerate cases.  In the general case, a lid is an
edge of the polygon's convex hull.  However, in degenerate cases where
three or more vertices are colinear, a lid can be a proper subset of a
convex hull edge according to Joss and Shannon's original definition
\cite{grunbaum-95}.  Although there are several different types of
degeneracies, only one type will actually affect our results.  We call
a pocket or flipturn \emph{degenerate} whenever the two edges just
outside the pocket lie on the same line.  In our illustrations of
flipturn sequences such as Figure \ref{Fig/sequence}, circled numbers
indicate degenerate flipturns.

Since flipturning about a proper subset of a convex hull edge may seem
unnatural, we offer the following alternative definition.  An
\emph{extended} pocket of a polygon is a chain of at least two edges
joining an adjacent pair of convex hull vertices.  An extended
flipturn rotates an extended pocket $180$ degrees about the midpoint
of its lid, which is a complete convex hull edge.  An extended pocket
or flipturn is \emph{degenerate} if and only if the two edges just
\emph{inside} the pocket lie on the same line.

Another alternative is proposed by Ahn \etal\ \cite{abchkm-fyl-00},
who define \emph{modified} pockets as follows.  Consider a standard
pocket with vertices $v_i, v_{i+1}, \dots, v_j$ (where index
arithmetic is modular).  If the nearby vertex $v_{j+1}$ lies on the
line through $v_i$ and $v_j$, then the chain of edges from $v_i$ to
$v_{j+1}$ is a modified pocket; otherwise, the standard pocket from
$v_i$ to $v_j$ is a modified pocket.  If the standard pocket is
degenerate, the modified pocket contains one of the two colinear
boundary edges.

\begin{figure}[htb]
\centerline{\footnotesize\sf
\begin{tabular}{c}\epsfig{file=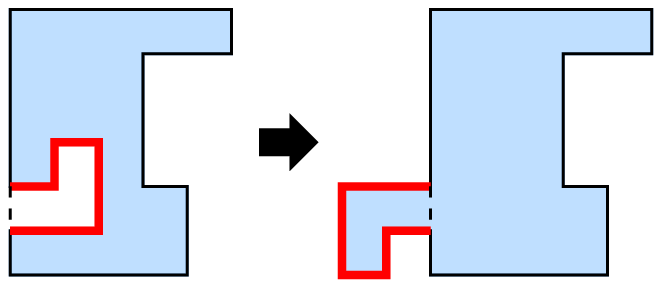,height=0.75in}\\(a)\end{tabular}
\hfil\hfil
\begin{tabular}{c}\epsfig{file=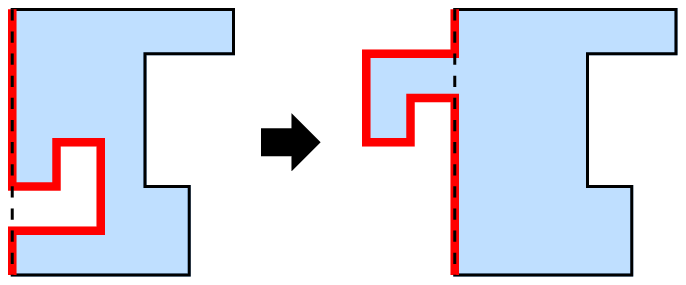,height=0.75in}\\(b)\end{tabular}
\hfil\hfil
\begin{tabular}{c}\epsfig{file=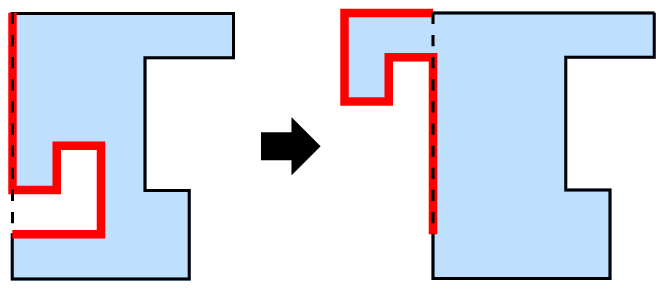,height=0.75in}\\(c)\end{tabular}
}
\caption{(a) A standard flipturn.  (b) An extended flipturn.  (c) A
modified flipturn.  Compare with Figure~\ref{Fig/1flipturn}.}
\label{Fig/1ortho}
\end{figure}

Figure~\ref{Fig/1ortho} illustrates a standard flipturn, an extended
flipturn, and one of two possible modified flipturn of the `same'
degenerate pocket of a polygon.  Note that a single extended flipturn
can simultaneously invert several standard or modified pockets.

In Section~\ref{ortho}, we will focus entirely on \emph{orthogonal}
polygons, each of whose edges is either horizontal or vertical.  We
say that a pocket or flipturn is \emph{orthogonal} if its lid is
horizontal or vertical and \emph{diagonal} otherwise.  In this
context, a pocket is degenerate if and only if it is orthogonal, and
so standard, extended, and modified orthogonal flipturns have
different behaviors, as Figure~\ref{Fig/1ortho} shows.  By any of our
three definitions, a diagonal flipturn reduces the number of vertices
of the polygon by two; specifically, the endpoints of the flipturned
pocket lie in the interior of edges of the new polygon.  If the input
polygon is in general position, \emph{every} flipturn will be
nondegenerate, and therefore diagonal.\footnote{We emphasize that
`general position' does not mean simply that no three vertices are
colinear.  In our context, an orthogonal polygon is in general
position if an arbitrary infinitesimal perturbation of its edge
lengths does not change the set of possible flipturn sequences.}
These observations immediately imply the following theorem.

\begin{theorem}
Exactly $(n-4)/2$ flipturns are necessary and sufficient to convexify
any orthogonal $n$-gon in general position, and these flipturns can be
chosen arbitrarily.
\end{theorem}

Thus, any discussion of flipturn sequences on orthogonal polygons only
becomes interesting if orthogonal flipturns are possible.  Even for
degenerate polygons, the fact that every diagonal flipturn removes two
vertices immediately implies the following upper and lower bounds.

\begin{theorem}
Any orthogonal $n$-gon is convexified by any sequence of $(n-4)/2$
diagonal flipturns.
\end{theorem}

\begin{theorem}
\label{trivlower}
At least $(n-4)/2$ flipturns are required to convexify any orthogonal
$n$-gon.
\end{theorem}

\emph{Every} modified flipturn on an orthogonal polygon removes two
vertices.  The somewhat convoluted definition of modified pockets
seems to have been developed precisely to avoid the `interesting'
consequences of degeneracies.  We immediately obtain the following
result, most of which is a special case of a theorem of Ahn
\etal~\cite{abchkm-fyl-00}.

\begin{theorem}
\label{mod}
Exactly $(n-4)/2$ modified flipturns are necessary and sufficient to
convexify any orthogonal $n$-gon, and these flipturns can be chosen
arbitrarily.
\end{theorem}


\section{Flipturn Sequences for Orthogonal Polygons}
\label{ortho}

In this section, we derive bounds on the maximum length of either the
shortest or longest convexifying flipturn sequences for orthogonal
polygons.  The bounds for the shortest sequence tell us how quickly we
can convexify a polygon if we choose flipturns intelligently; the
longest sequence bounds tell us how many flipturns we can perform even
if we choose flipturns blindly.  Our results are summarized in the
first two rows of Table \ref{results}.  Since Theorem \ref{mod}
completely characterizes the lengths of modified flipturn sequences
for orthogonal polygons, this section will focus entirely on standard
and extended flipturns.

\subsection{Order Matters}

\begin{figure}[htb]
\centerline{\epsfig{file=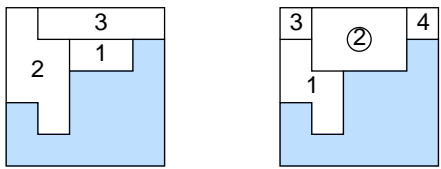,height=0.85in}}
\caption{The same polygon can have standard or extended flipturn
sequences of different lengths.}
\label{Fig/diff}
\end{figure}

Once we recognize the possibility of orthogonal flipturns, it is easy
to construct polygons such as in Figure~\ref{Fig/diff} that have
flipturn sequences of different lengths.  The polygon has two pockets;
flipturning one of them first creates an orthogonal pocket, and
flipturning the other first does not.

In fact, as the following theorem shows, the shortest and longest
flipturn sequences may differ significantly in length.

\begin{theorem}
For infinitely many $n$, there is an orthogonal $n$-gon whose shortest
and longest standard or extended flipturn sequences differ in length
by at least $(n-4)/4$.
\end{theorem}

\begin{figure}[htb]
\centerline{\epsfig{file=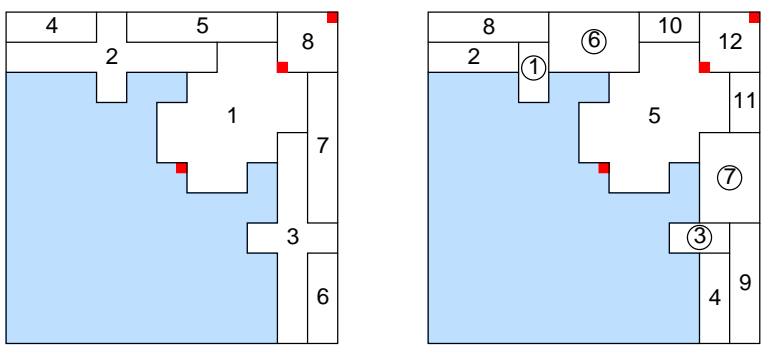,height=1.75in}}
\caption{An orthogonal polygon that can be convexified with either
$(n-4)/2$ or $\floor{3(n-4)/4}$ flipturns.  The small squares contain
a recursive copy of the polygon.}
\label{Fig/delta}
\end{figure}

\begin{proof}
Figure~\ref{Fig/delta} illustrates the recursive construction of such
a polygon, for all $n$ of the form $16k+4$.  The shortest flipturn
sequence for the polygon includes only diagonal flipturns and
therefore has length $(n-4)/2$.  Another sequence, which we believe to
be the longest, requires twelve flipturns to remove every $16$
vertices.  Figure~\ref{Fig/delta} illustrates this long sequence of
standard flipturns; the corresponding extended flipturn sequence is
essentially equivalent.
\end{proof}

\subsection{Well-chosen Flipturns}
\label{shortest}

Here we develop upper and lower bounds on the length of the shortest
sequence of flipturns required to convexify an orthogonal polygon.
For any polygon $P$, let $\Box(P)$ denote its axis-aligned bounding
rectangle.

\begin{theorem}
\label{shortlow}
For all $n$, there is an orthogonal $n$-gon that requires
$\floor{3(n-4)/4}$ standard or extended flipturns to convexify.
\end{theorem}

\begin{figure}[htb]
\centerline{\epsfig{file=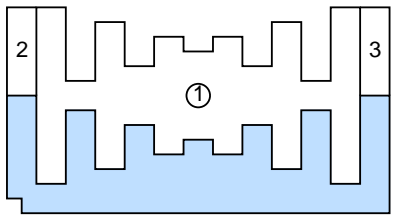,height=1.1in}}
\caption{An orthogonal $n$-gon requiring $\floor{3(n-4)/4}$ flipturns
to convexify.}
\label{Fig/3fourths}
\end{figure}

\begin{proof}
When $n$ is a multiple of $4$, the polygon consists of a horizontally
symmetric rectangular `comb' with $n/4$ `teeth'; if $n$ is not a
multiple of $4$, we add a small rectangular notch in a bottom corner
of the polygon.  See Figure~\ref{Fig/3fourths}.  (We consider a
rectangle to be a comb with one tooth.)  Both the teeth and the gaps
between them decrease in height as they approach the middle of the
polygon.  Since the polygon is symmetric about its vertical bisecting
line, standard and extended flipturns have exactly the same effect.
The only way to eliminate the comb is through a sequence of orthogonal
flipturns across the top edge of the polygon's bounding box; each such
flipturn eliminates exactly one tooth.  It easily follows that
\emph{every} flipturn sequence for this polygon has length
$\floor{3(n-4)/4}$.
\end{proof}

\begin{lemma}
\label{corner}
Let $P$ be an orthogonal polygon.
\begin{enumerate}\cramped
\item[\textup{\textbf{(a)}}]
If some vertex of $\Box(P)$ is not a vertex of $P$, then $P$ has a
diagonal pocket.
\item[\textup{\textbf{(b)}}]
If two adjacent vertices of $\Box(P)$ are not vertices of $P$, then we
can perform at least two consecutive diagonal flipturns on $P$.
\end{enumerate}
\end{lemma}

\begin{figure}[htb]
\centerline{\epsfig{file=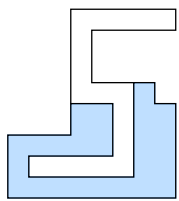,height=1in}}
\caption{Flipturning one diagonal pocket can hide another one.}
\label{Fig/corners}
\end{figure}

\begin{proof}
\textbf{(a)} Suppose some corner of $\Box(P)$ is not a vertex of $P$.
Some edge of $\conv(P)$ lies on a line separating the missing corner
from the interior of $P$.  This edge contains a diagonal lid.

\textbf{(b)} Without loss of generality, suppose $P$ does not contain
the top left and top right vertices of $\Box(P)$.  Part (a) implies
that $P$ has at least two diagonal pockets.  Let~$Q$ be the result of
flipturning one of these pockets.  Since the width of the flipturned
pocket is less than the width of $P$, and thus less than the width
of~$Q$, at least one of the upper corners of $\Box(Q)$ is not a vertex
of~$Q$.  (As Figure~\ref{Fig/corners} shows, flipturning one pocket
can capture the opposite corner.)  Thus, by part~(a), $Q$~still has at
least one diagonal pocket.
\end{proof}

This lemma is a special case of a more general result, whose proof we
omit: If any $k$ vertices of $\Box(P)$ are not vertices of $P$, then
we can perform at least $k$ consecutive diagonal flipturns on $P$.

\begin{theorem}
\label{extupper}
Any orthogonal $n$-gon can be convexified by a sequence of at most
$\floor{3(n-4)/4}$ extended flipturns.
\end{theorem}

\begin{proof}
We achieve the stated upper bound by performing an orthogonal extended
flipturn only when no diagonal pockets are available.  By
Lemma~\ref{corner}, we are forced to perform an orthogonal flipturn on
a polygon $P$ if and only if all four corners of $\Box(P)$ are also
vertices of~$P$.

Let $P$ be a nonconvex orthogonal $n$-gon with bounding box $\Box(P)$,
and suppose $P$ has no diagonal pockets.  Without loss of generality,
suppose $P$ has an extended orthogonal pocket whose lid $lr$ is the
top edge of $\Box(P)$.  This pocket obviously lies strictly between
the vertical lines through $l$ and $r$.  Let $P_1$ be the polygon that
results when this extended pocket is flipturned.  The highest vertices
of $P_1$ are vertices of the newly flipturned pocket, and thus must
lie strictly between the vertical lines through $l$ and~$r$.  Thus,
neither of the top vertices of $\Box(P_1)$ is a vertex of $P_1$, and
by Lemma~\ref{corner}, we can perform at least two consecutive
diagonal flipturns on $P_1$.  See Figure~\ref{Fig/ext}.

\begin{figure}[htb]
\centerline{\epsfig{file=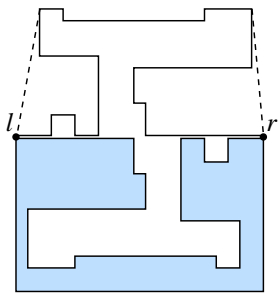,height=1.5in}}
\caption{Any orthogonal extended flipturn creates at least two
diagonal pockets.}
\label{Fig/ext}
\end{figure}

In other words, any orthogonal extended flipturn can be followed by at
least two diagonal flipturns.  Thus, if we perform orthogonal
flipturns only when no diagonal flipturn is available, any three
consecutive flipturns eliminate at least four vertices.
\end{proof}

Theorem \ref{shortlow} implies that this result is the best possible
for extended flipturns.  For standard flipturns, we obtain the
following slightly weaker upper bound.

\begin{theorem}
\label{upper}
Any orthogonal $n$-gon can be convexified by a sequence of at most
$\floor{5(n-4)/6}$ standard flipturns.
\end{theorem}

\begin{proof}
As in the previous theorem, we achieve the upper bound by performing
orthogonal flipturns only when no diagonal flipturn is available.
However, we also choose orthogonal flipturns carefully if more than
one is available.  Say that an orthogonal flipturn is \emph{good} if
it can be followed by at least two diagonal flipturns and \emph{bad}
otherwise.  We will perform a bad orthogonal flipturn only if no good
orthogonal flipturn or diagonal flipturn is available.

Let $P$ be an orthogonal polygon.  Without loss of generality,
consider a forced orthogonal flipturn whose lid $bc$ lies on the top
edge of $\Box(P)$, and let $P_1$ be the polygon resulting from this
flipturn.  See Figure~\ref{Fig/forced}(a).  The lid endpoints $b$ and
$c$ must lie in two different pockets of $P_1$, since the flipturned
pocket touches the top of $\Box(P_1)$.  The horizontal width of the
pocket must be less than the horizontal width of $P$, so $P_1$ cannot
have both the upper left and upper right corners of $\Box(P_1)$ as
vertices.  Thus, by Lemma \ref{corner}, any forced orthogonal flipturn
can be followed first by a diagonal flipturn and then by at least one
more (possibly orthogonal) flipturn.  In particular, any bad flipturn
can be followed by exactly one diagonal flipturn.

\begin{figure}[htb]
\centerline{\footnotesize\sf
    \begin{tabular}{c@{\qquad}c@{\qquad}c}
	\epsfig{file=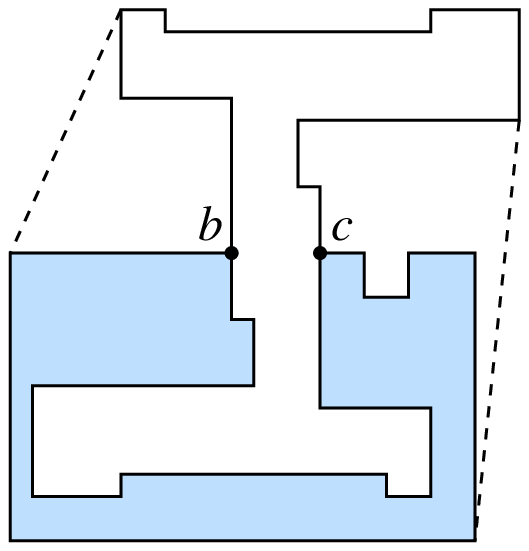,height=1.5in} &
	\epsfig{file=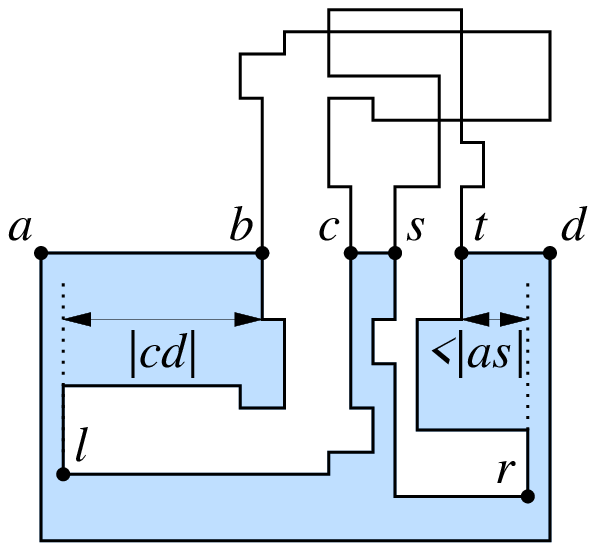,height=1.5in} &
	\epsfig{file=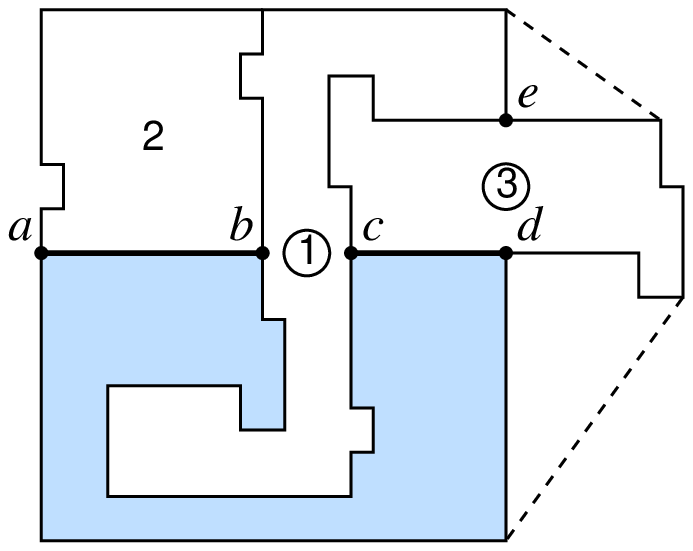,height=1.5in} \\
	 (a) & (b) & (c)
    \end{tabular}}
\caption{(a) A forced orthogonal flipturn creates at least two
pockets, at least one of which is diagonal.  (b) A polygon with only
bad pockets cannot have both dexter and sinister pockets on the same
edge.  (c) A forced bad orthogonal flipturn
(flipturn~{\protect\textcircled{\protect\scriptsize 1}})
creates a good orthogonal pocket
(flipturn~{\protect\textcircled{\protect\scriptsize 3}}).}
\label{Fig/forced}
\end{figure}

Let $P$ be a polygon with no diagonal pockets or good orthogonal
pockets.  Consider a bad orthogonal flipturn whose lid $bc$ is a
subset of the top edge $ad$ of $\Box(P)$, and let $P_1$ be the
resulting polygon.  Exactly one of the top corners of $\Box(P_1)$ is a
vertex of $P_1$.  If this is the top right corner, call pocket $bc$
\emph{dexter}; otherwise, call it \emph{sinister}.  Without loss of
generality, suppose the pocket $bc$ is dexter.  Let $P_2$ be the
polygon resulting from the only available diagonal flipturn, whose lid
is the upper left edge of $\conv(P_1)$.  Since $P_2$ must have no
diagonal pockets, this flipturn moves vertex $b$ to the upper left
corner of $\Box(P_2)$.  See Figure~\ref{Fig/forced}(c).

If some pocket had a lid in $ab$, that pocket would be inverted by the
diagonal flipturn on $P_1$ and~$P_2$ would have a diagonal pocket,
contradicting our assumption that pocket $bc$ is bad.  Similarly, if
there is a bad pocket with lid in $cd$, it cannot be dexter.  Suppose
there is a sinister pocket with lid $st\subset cd$.  Let $l$ be a
leftmost point in pocket $bc$, and let $r$ be a rightmost point in
pocket $st$.  See Figure~\ref{Fig/forced}(b).  The horizontal distance
from $l$ to $b$ must be equal to $\abs{cd}$, and the horizontal
distance from $t$ to $r$ must equal to $\abs{as}$, since both pockets
are bad.  But this is impossible, since $\abs{cd}+\abs{as} >
\abs{ad}$.  We conclude that $bc$ must be the \emph{only} lid on the
top edge of $\Box(P)$.

Now consider the orthogonal pocket of $P_2$ created when pocket $bc$
is flipturned.  Its lid $de$ lies on the right edge of $\Box(P_2)$.
We claim that this pocket must be good.  Let $P_3$ be the resulting
polygon when this pocket is flipturned.  Since $cd$ is the bottommost
edge of pocket $de$, nothing in $P_3$ lies above and to the right of
vertex $e$, so the upper right vertex of $\Box(P_3)$ is not a vertex
of $P_3$.  Since the height of pocket $de$ is less than the height of
the original polygon $P$, the bottom right vertex of $\Box(P_3)$ is
also not a vertex of $P_3$.  Therefore, by Lemma \ref{corner}, $P_3$
can undergo at least two consecutive flipturns.

We have just shown that any forced bad flipturn is immediately
followed by a diagonal flipturn, a good orthogonal flipturn, and then
two diagonal flipturns.  Thus, any sequence of five consecutive
flipturns contains at least three diagonal flipturns, which remove at
least six vertices from the polygon.
\end{proof}

We do not believe that this upper bound is tight.  In the following
section, we will show that the algorithm used to prove the upper bound
may not produce the shortest flipturn sequence.

\subsection{Arbitrary Flipturns}
\label{longest}

In this section, we consider the length of the \emph{longest} sequence
of flipturns that an orthogonal polygon can undergo.

\begin{theorem}
\label{bracket}
For all $n>4$, the longest standard or extended flipturn sequence
for any orthogonal $n$-gon has length at most $n-5$.
\end{theorem}

\begin{proof}
We call an edge of an orthogonal polygon a \emph{bracket} if both its
vertices are convex or both its vertices are concave.  An orthogonal
$n$-gon has at least four brackets (the highest, leftmost, lowest, and
rightmost edges) and at most $n-2$ brackets.

We claim that flipturns do not increase the number of brackets, and
that any orthogonal flipturn decreases the number of brackets by two.
Let $P$ be an orthogonal polygon and let $Q$ be the result of one
flipturn.  Any bracket of $P$ that lies completely outside the
flipturned pocket is still a bracket in $Q$; any bracket completely
inside the flipturned pocket is inverted, but remains a bracket.
Thus, to prove our claim, it suffices to consider just four edges,
namely, the two edges adjacent to each endpoint of the lid.  After
symmetry considerations, there are only three cases to check for
orthogonal pockets and ten cases for diagonal pockets.  These cases
are illustrated in Figure~\ref{Fig/bracket}.

\begin{figure}[htb]
\centerline{\epsfig{file=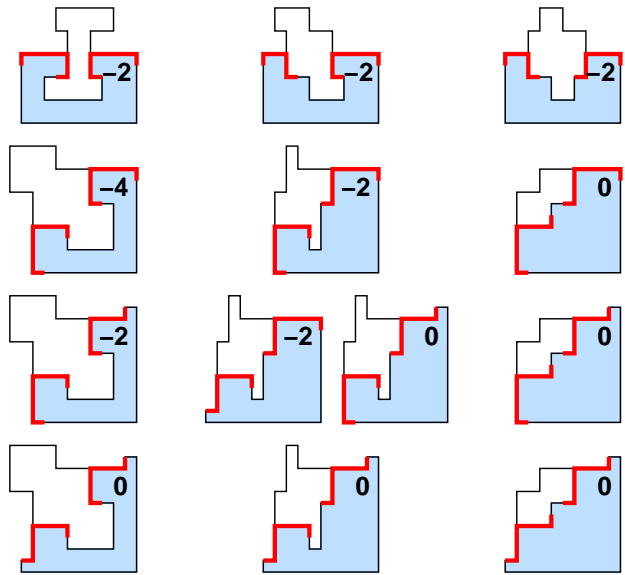,height=3in}}
\caption{Thirteen classes of flipturns and the number of brackets they
remove.  Only the bold (red) edges are important.  The top row shows
orthogonal flipturns; the other rows show diagonal flipturns with two,
one, and no outer brackets.  The columns show flipturns with two, one,
and no inner brackets.  Symmetric cases are omitted.  Compare
with Figure~\protect\ref{Fig/bracket2}.}
\label{Fig/bracket}
\end{figure}

Since each orthogonal flipturn removes two brackets, and no diagonal
flipturn adds brackets, there can be at most $(n-6)/2$ orthogonal
flipturns.  Since each diagonal flipturn removes two vertices, and no
orthogonal flipturn adds vertices there can be at most $(n-4)/2$
diagonal flipturns.  Thus, there can be at most $(n-6+n-4)/2 = n-5$
flipturns altogether.
\end{proof}

We can improve this upper bound slightly in the special case of
\emph{lattice} polygons---orthogonal polygons where every edge has
unit length (or more generally, where every edge has integer length
and $n$ denotes the perimeter instead of the number of edges).

\begin{theorem}
The longest flipturn sequence for any lattice $n$-gon has length at
most $n-2\sqrt{n}$.
\end{theorem}

\begin{proof}
In any convexifying sequence, there are exactly $n/2 - 2$ diagonal
flipturns.  Every orthogonal flipturn increases the perimeter of the
polygon's bounding box by at least $2$.  The initial bounding box has
perimeter at least $4(\sqrt{n}-1)$, and the final rectangle has
perimeter exactly $n$, so the maximum number of orthogonal flipturns
is at most $n/2 - 2\sqrt{n} + 2$.
\end{proof}

How tight is the $n-5$ upper bound?  As in the case of the shortest
flipturn sequence, the answer depends on whether we consider standard
or extended flipturns.  Unfortunately, we do not obtain an exact
answer in either case.

\begin{theorem}
For all $n$, there is an orthogonal $n$-gon that can undergo
$\floor{3(n-4)/4}$ extended flipturns.
\end{theorem}

\begin{proof}
This follows directly from Theorem \ref{shortlow}.
\end{proof}

\begin{theorem}
For all $n$, there is an orthogonal $n$-gon that can undergo
$\floor{5(n-4)/6}$ standard flipturns.
\end{theorem}

\begin{figure}[htb]
\centerline{\epsfig{file=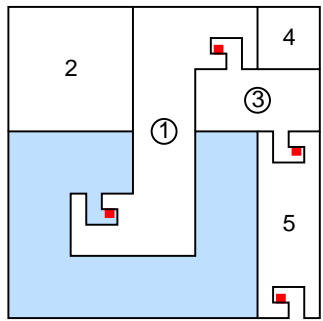,height=1.65in}}
\caption{An orthogonal $n$-gon that can undergo $\ceil{5(n-4)/6}$
standard flipturns.  Two levels of recursion are shown.  The small
squares contain a recursive copy of the polygon.}
\label{Fig/5sixths}
\end{figure}

\begin{proof}
We construct an orthogonal $n$-gon $P_n$ essentially by following the
proof of Theorem \ref{upper}.  $P_4$ is a rectangle.  $P_6$ is an
\textbf{L}-shaped hexagon, which is convexified by one flipturn.
$P_8$ is a rectangle with a rectangular orthogonal pocket in one side,
which requires three flipturns to convexify.  For all $n\ge 10$, $P_n$
consists of a rectangle with a single \textbf{L}-shaped pocket, where
the tail of the \textbf{L} is an inverted and reflected copy of
$P_{n-6}$.  See Figure~\ref{Fig/5sixths}.  In the language of the
proof of Theorem~\ref{upper}, $P_n$'s only pocket is
\emph{bad}---flipturning it creates one diagonal pocket and one
orthogonal pocket.  If we flipturn diagonal pockets whenever possible,
the first five flipturns eliminate six vertices and leave a distorted
$P_{n-6}$.  The theorem follows by induction.
\end{proof}

\begin{figure}[htb]
\centerline{\footnotesize\sf
    \begin{tabular}{c@{\qquad\qquad}c}
        \epsfig{file=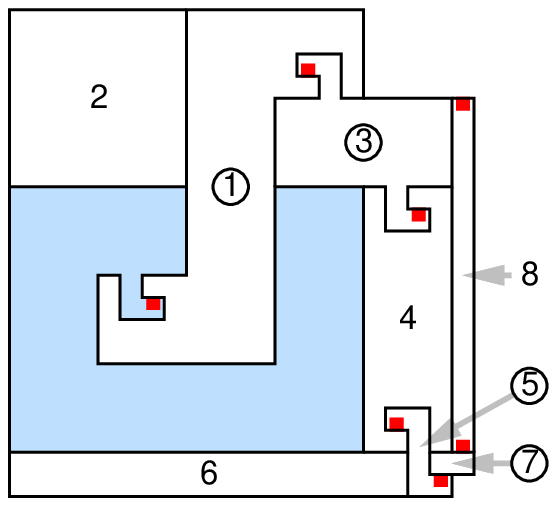,height=1.75in} &
        \epsfig{file=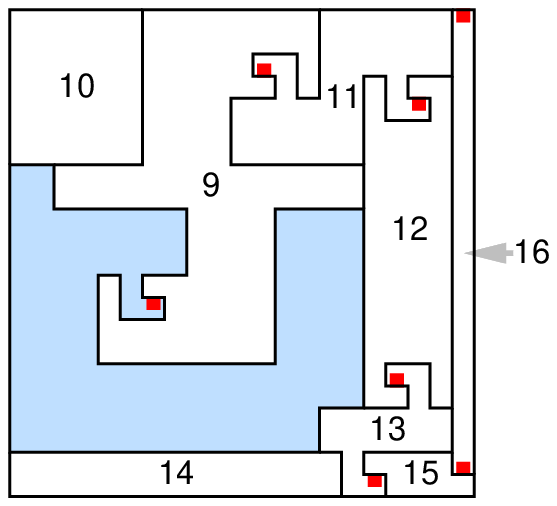,height=1.75in} \\
	(a)\quad~ & (b)\quad~
    \end{tabular}}
\caption{The algorithm from Theorem~\ref{upper} is not optimal.
(a)~The first eight flipturns in a shorter convexifying sequence.
(b)~The next eight flipturns; the polygon has been distorted to
emphasize relevant details.}
\label{Fig/semigreed}
\end{figure}

To prove Theorem \ref{upper}, we used an algorithm that always prefers
diagonal flipturns to orthogonal flipturns and good orthogonal
flipturns to bad orthogonal flipturns.  We can use the polygon $P_n$
from the previous proof to show that this algorithm is not optimal, by
demonstrating a shorter convexifying flipturn sequence.
Figure~\ref{Fig/semigreed} shows the first sixteen flipturns performed
by a modified algorithm, which ignores diagonal `notches' in the upper
and lower right corners of the polygon.  Figure~\ref{Fig/semigreed}(b)
is distorted to reveal relevant but otherwise invisible details; the
distortion does not change which flipturns we can perform at any time.
These $16$ flipturns remove $24$ vertices, thereby transforming $P_n$
into $P_{n-24}$.  Thus, by induction, we can convexify $P_n$ in only
$2(n-4)/3$ flipturns whenever $n-4$ is a multiple of $24$.

The ignored `notches' are precisely the diagonal flipturns that do not
remove brackets; see Theorem \ref{bracket}.  Perhaps a modified
algorithm that tries to reduce the number of brackets as fast as
possible, as well as the number of vertices, would improve Theorem
\ref{upper}.  We leave the development of such an algorithm as an
intriguing open problem.

Finally, we observe that $P_n$ can be convexified using exactly
$\ceil{2(n-3)/3}$ extended flipturns.  We leave the proof as an easy
exercise of the reader.


\section{Flipturn Sequences for Arbitrary Polygons}
\label{general}

In this section, we derive upper bounds for the longest flipturn
sequences of arbitrary polygons, generalizing both our earlier results
for orthogonal polygons and the modified flipturn results of Ahn
\etal\ \cite{abchkm-fyl-00}.

Consider an arbitrary polygon $P$ whose boundary is oriented
counterclockwise.  Let $\vec{e}$ denote the direction of any
(oriented) edge $e$ in $P$, let~$S$ be the set of all such edge
directions and their edge reversals.  We clearly have $s \le \abs{S}
\le 2s$, where $s$ is the number of distinct edge slopes.  Ahn \etal\
define the \emph{discrete angle} at a vertex $v = e\cap e'$ to be one
more than the number of elements of $S$ inside the angle between
$\vec{e}$ and $\vec{e}'$.  The \emph{total discrete angle} $D(P)$ is
the sum of the discrete angles at the vertices of~$P$.

Ahn \etal\ prove the following lemma~\cite{abchkm-fyl-00}.  (Only the
first half of this lemma is stated explicitly, but their proof implies
the second half as well.)

\begin{lemma}
\label{angles}
Every non-degenerate flipturn decreases $D(P)$ by at least two, and
every degenerate flipturn leaves $D(P)$ unchanged.
\end{lemma}

Ahn \etal\ also prove that $D(P) \le n(s-1)$ in general and $D(P) =
2s$ if $P$~is convex.  Thus, Lemma \ref{angles} immediately implies
that $\ceil{(ns-n-2s)/2} \le n(n-3)/2$ nondegenerate flipturns suffice
to convexify any polygon.  However, since no bound was previously
known for the number of degenerate flipturns, this bound does not
apply to degenerate polygons.  To avoid this problem, Ahn \etal\
introduce modified flipturns, for which degeneracies do not exist.  To
account for degenerate flipturns under the standard definition, we
study the change in the number of \emph{brackets}, here denoted
$B(P)$, as in Section~\ref{longest}.

\begin{lemma}
\label{L:bracket2}
Every non-degenerate standard or extended flipturn increases $B(P)$
by at most two, and every degenerate standard or extended flipturn
decreases $B(P)$ by at least two.
\end{lemma}

\begin{figure}[htb]
\centerline{\footnotesize\sf
    \begin{tabular}{c}
	\epsfig{file=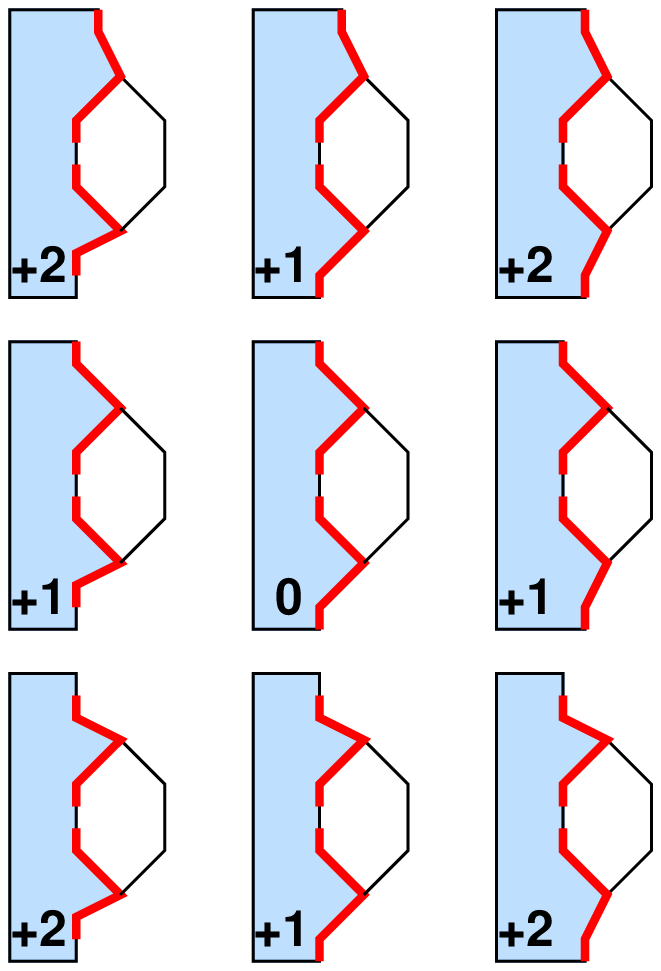,height=2.5in}\\$b=0$
    \end{tabular}
\hfil\hfil
    \begin{tabular}{c}
	\epsfig{file=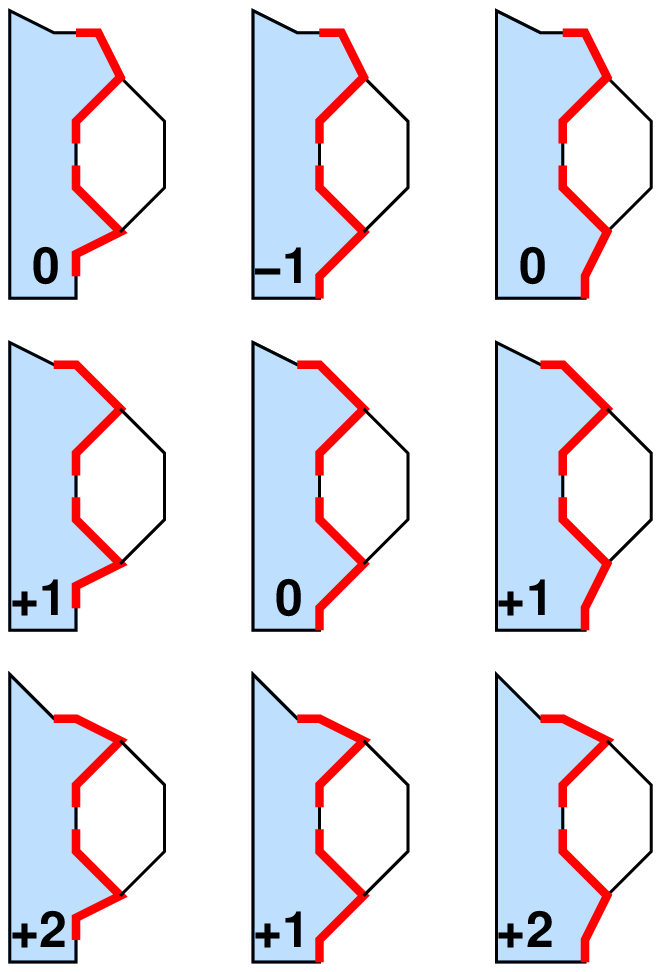,height=2.5in}\\$b=1$, outer
    \end{tabular}
\hfil\hfil
    \begin{tabular}{c}
	\epsfig{file=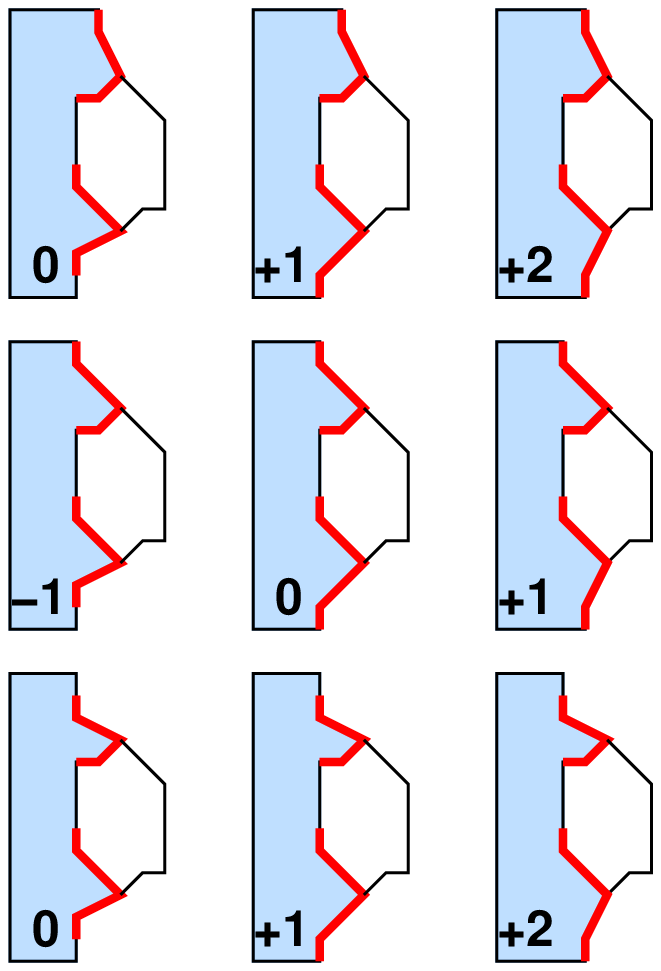,height=2.5in}\\$b=1$, inner
    \end{tabular}}
\caption{Twenty-seven classes of nondegenerate flipturns and the
number of brackets they add or remove.  Only the bold (red) edges are
important.  Symmetric cases are omitted.  Compare with
Figure~\protect\ref{Fig/bracket}.}
\label{Fig/bracket2}
\end{figure}

\begin{proof}
Let $P$ be a simple polygon and let $P'$ be the result of one
flipturn.  As we argued in the proof of Theorem~\ref{bracket}, it
suffices to focus on brackets touching the endpoints of the lid.  Let
$b$ and~$b'$ denote the number of boundary brackets in $P$ and $P'$,
respectively, so that $B(P') = B(P) - b + b'$.

For nondegenerate flipturns, it suffices to consider only flipturns
with $b\le 1$, since $b'$ is never more than $4$.  There are three
cases to consider: no boundary brackets, one outer boundary bracket,
and one inner boundary bracket.  For each of these, there are nine
subcases, depending on whether each lid endpoint becomes a convex
vertex, becomes a concave vertex, or disappears after the flipturn.
These cases are illustrated in Figure~\ref{Fig/bracket2}.

Standard degenerate flipturns always have two outer brackets, and both
lid endpoints always become concave vertices.  Thus, there are only
three cases to consider, depending on the number of inner brackets,
exactly as in Theorem~\ref{bracket}.  Similar arguments apply to
extended flipturns.
\end{proof}

\begin{theorem}
\label{bracket2}
Every $s$-oriented polygon is convexified after any sequence of
$ns-\floor{(n+5s)/2}-1$ standard or extended flipturns.
\end{theorem}

\begin{proof}
We define the \emph{potential} $\Phi$ of a polygon to be its discrete
angle plus half the number of brackets, that is, $\Phi = D + B/2$.
For the initial polygon $P$, we have $D \le n(s-1)$ and $B\leq n-2$,
so the initial potential $\Phi$ is at most $ns-n/2-1$.  For the final
convex polygon, we have $D=2s$ and $B\ge s$, so the final potential
$\Phi^*$ is at least $5s/2$.  By Lemmas \ref{angles} and
\ref{L:bracket2}, every flipturn reduces the potential by at least
one.  Thus, after any sequence of $\ceil{\Phi^* - \Phi} = \ceil{ns -
n/2 - 5s/2 - 1}$ flipturns, the polygon must be convex.
\end{proof}

If we set $s=n$, we obtain the upper bound $n^2-3n-1$ for arbitrary
simple polygons.  However, if $s=n$, there can be no degenerate
flipturns, so the discrete angle results from Ahn \etal\ apply
directly, giving us the upper bound $n(n-3)/2$.  Hence, the actual
worst case arises when $s=n-1$.

\begin{corollary}
Every simple polygon is convexified after any sequence of $n^2-4n+2$
standard or extended flipturns.
\end{corollary}

This upper bound is almost certainly not tight.  Intuitively, if $s$
large, only a few pairs of edges can have the same slope, so the
maximum number of degenerate flipturns should be small.

We can improve our results in some cases using a different definition
of discrete angle.  Let $T$ denote the set of edge directions
(\emph{without} their reversals), let $t = \abs{T}$, and let $h \le
t-1$ be the maximum number of edge directions that fit in an open
half-circle.  The discrete angle at any vertex is at most $h-1$, so
$D(P) \le n(h-1) \le n(t-2)$ for any polygon $P$; if $P$ is convex,
then $D(P) = t$.  Lemma \ref{angles} still holds under this new
definition.  Thus, we obtain the following upper bounds.

\begin{theorem}
\label{half}
Every simple polygon is convexified after any sequence of ${\ceil{(nh
- n - t)/2}} \le {\ceil{t(n-1)/2}-n}$ modified flipturns or $nh -
\floor{(n+3t)/2} - 1 \le nt - \floor{3(n+t)/2} - 1$ standard or
extended flipturns.
\end{theorem}

This theorem improves all earlier results whenever $h$ is
significantly smaller than~$t$.  For general polygons, setting $h =
t-1 = n-1$ gives us the same $n(n-3)/2$ upper bound for modified
flipturns.  For standard or extended flipturns, however, we obtain a
very slight improvement by setting $h = t-1 = n-2$.

\begin{corollary}
Every simple polygon is convexified after any sequence of $n^2-4n+1$
standard or extended flipturns.
\end{corollary}

We close this section with some obvious open questions.
Asymptotically, our bounds agree with Joss and Shannon's
conjecture~\cite{grunbaum-95}---any polygon can indeed be convexified
by $O(n^2)$ flipturns---but there is still a significant gap between
our upper bounds and the $(n-2)^2/4$ lower bound construction of
Biedl~\cite{biedl-2000}.  We, like Joss and Shannon, conjecture that
the correct answer is closer to $n^2/4$.

A more interesting open question concerns the length of
\emph{shortest} flipturn sequences for general polygons.  The best
lower bounds are those derived for orthogonal polygons in Section
\ref{ortho}, but not subquadratic upper bounds are known.  Can
arbitrary polygons be convexified with only $O(n)$ flipturns, or does
some polygon require a super-linear number of flipturns to convexify?


\section{Data Structures for Flipturns}
\label{algo}

In this section, we describe efficient data structures for executing a
sequence of flipturns on any simple (not necessarily orthogonal)
polygon.  We first describe a simpler data structure that maintains an
implicit description of a polygon $P$ as flipturns are performed,
without worrying about how the flipturns are chosen.  Then we will
describe how to maintain the convex hull of $P$ as we perform
flipturns, so that we can determine which flipturns are available at
any time.

\begin{lemma}
\label{reverse}
After $O(n)$ preprocessing time, we can maintain an implicit
description of a simple $n$-gon in $O(\log n)$ time per flipturn,
using a data structure of size $O(n)$.
\end{lemma}

\begin{proof}
It suffices to store only the slopes and lengths of the edges in the
proper order, without explicitly storing the vertex coordinates.  Any
flipturn reverses a contiguous chain of edges, namely, the edges of
the flipturned pocket.  Our goal, therefore, is to maintain a circular
list of items subject to the operation $\mathsc{Reverse}(s,t)$, which
reverses the sublist starting with item $s$ and ending with item $t$.
For example, if the initial list is $(a,b,c,d,e,f,g,h)$, then
$\mathsc{Reverse}(c,f)$ produces the list $(a,b,f,e,d,c,g,h)$, after
which $\mathsc{Reverse}(d,a)$ produces $(h,g,c,d,b,f,e,a)$.

We store the edges in the leaves of a balanced binary search tree,
initially in counterclockwise order around the polygon.  Rather than
explicitly reversing chains of edges, we will store a \emph{reversal
bit}~$r_v$ at every node~$v$, indicating whether that subtree should
be considered reversed, relative to the orientation of the subtree
rooted at $v$'s parent.  Initially, all reversal bits are set to~$0$.

\begin{figure}[htb]
\centerline{\epsfig{file=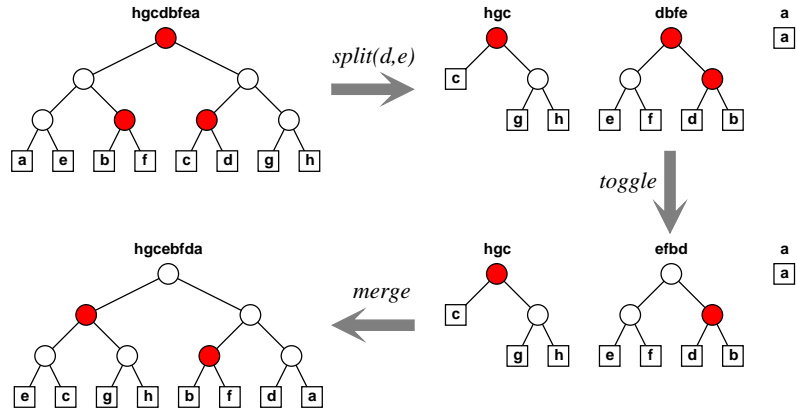,height=2.125in}}
\caption{Executing $\mathsc{Reverse}(d,e)$ to transform
$(h,g,c,d,b,f,e,a)$ into $(h,g,c,e,f,b,d,a)$.  Solid nodes have
reversal bits set to $1$.}
\label{Fig/reverse}
\end{figure}

Our algorithm for $\mathsc{Reverse}(s,t)$ is illustrated in
Figure~\ref{Fig/reverse}.  Without loss of generality, we assume that
$s$ appears before~$t$ in the linear order stored in the tree;
otherwise, we simply toggle the reversal bit at the root and call
$\mathsc{Reverse}(t,s)$.  First, we split the tree into three
subtrees, containing the items to the left of~$s$, items between $s$
and $t$, and the items to the right of~$t$.  Second, we toggle the
reversal bit at the root of the middle tree.  Finally, we merge the
three trees back together.  Each split or merge can be performed using
$O(\log n)$ rotations (using red-black trees \cite{gs-dfbt-78}, splay
trees \cite{st-sabst-85}, or treaps \cite{sa-rst-96}, for example),
and we can easily propagate the reversal bits correctly at each
rotation.
\end{proof}

To maintain the convex hull of a polygon under flipturns, we use a
variant of the dynamic convex hull data structure of Overmars and van
Leeuwen \cite{ol-mcp-81}, which maintains the convex hull of a
dynamically changing set of points in $O(\log^2 n)$ time per insertion
or deletion.  Their data structure consists of a balanced binary tree
that allows insertions, deletions, splits, and merges in $O(\log n)$
time.  The points are stored at the leaves of this tree, ordered by
their $x$-coordinates.  Each node in the tree stores the convex hull
of the points in its subtree; we call this the node's \emph{subhull}.
Except at the root, these subhulls are not stored explicitly; rather,
each node stores only the chain of edges of its subhull that are not
in its parent's subhull.  Overmars and van Leeuwen show that any
node's subhull can be computed in $O(\log n)$ time from its children's
subhulls, by finding the outer common tangent lines.

There are several differences between our problem and the standard
dynamic convex hull problem.  The most significant difference is that
we need to support an operation similar to $\mathsc{Reverse}$ in
polylogarithmic time.  This requires us to store the vertices in their
order of appearance around the polygon, rather than in any coordinate
order.  Moreover, since a linear number of vertices could be affected
by a flipturn, our data structure must implicitly represent both the
order and the locations of the vertices.  A second significant
difference lies in the structure of the subhulls.  In Overmars and van
Leeuwen's data structure, the subhulls of any two siblings in the tree
are separated by a known vertical line.  In our case, sibling subhulls
are \emph{pseudo-disks}: either they have disjoint interiors, or they
have nested closures, or their boundaries intersect transversely at
exactly two points.  Distinguishing these three cases and merging the
subhulls in each case requires considerably more effort.  Finally, one
minor difference is that for standard flipturns, we must maintain the
complete sequence of polygon vertices on the boundary of the convex
hull, not just the convex hull vertices.  This requires only trivial
modifications, which do not deserve further mention.

\begin{lemma}
\label{convex}
After $O(n\log n)$ preprocessing time, we can maintain an implicit
description of the convex hull of a simple $n$-gon in $O(\log^4 n)$
time per flipturn, using a data structure of size $O(n)$.
\end{lemma}

\begin{proof}
We maintain the polygon vertices in a balanced binary tree, similarly
to the proof of Lemma~\ref{reverse}.  The coordinates of the points
are represented implicitly by storing a triple $(r_v, x_v, y_v)$ at
each internal node $v$, encoding an affine transformation to be
applied to all edges in the subtree of $v$.  Specifically, $(x_v,
y_v)$ is a translation vector for all edges in $v$'s subtree if $r_v =
0$ and a point of reflection if $r_v=1$.  Initially, $r_v = x_v = y_v
= 0$ for all nodes $v$.  The actual position of any vertex can be
recovered in $O(\log n)$ time by composing the transformations along
the path up to the root.  We can easily maintain these triples under
rotations, splits, and merges, similarly to the $\mathsc{Reverse}$
algorithm described earlier.  We omit the unenlightening details.

Each node in this tree also stores the portion of its subhull not
included in its parent's subhull.  Specifically, we store the vertices
of this convex chain in a secondary balanced binary tree.  Instead of
explicitly storing the coordinates of the vertices of this chain,
however, we store only pointers to the appropriate leaves in the
primary binary tree.  The coordinates of any point can be recovered in
$O(\log n)$ time by composing the linear transformations stored on the
path up from the point's leaf.

It remains only to show that we can merge any two sibling subhulls
quickly.  If we can merge two sibling subhulls in time $T(n)$ when all
vertex coordinates are given explicitly, then we can update the convex
hull of $P$ in time $O(T(n)\log^2 n)$ per flipturn.  One logarithmic
factor is the number of merges we must perform for each flipturn; the
other is the cost of accessing the implicitly-stored vertex
coordinates.

Let $C$ be the chain of polygon edges associated with some node $v$ in
the primary binary tree, and let $A$ and $B$ be the subchains
associated with the left and right children of $v$, respectively.
Since $C$ has no self-intersections, the boundaries of the convex
hulls $\conv(A)$ and $\conv(B)$ can intersect in at most two points.
If the hull boundaries do not intersect, then the hulls can be either
disjoint or nested.  See Figure~\ref{Fig/subchains}.

\begin{figure}[htb]
\centering\footnotesize\sf
\begin{tabular}{c}\epsfig{file=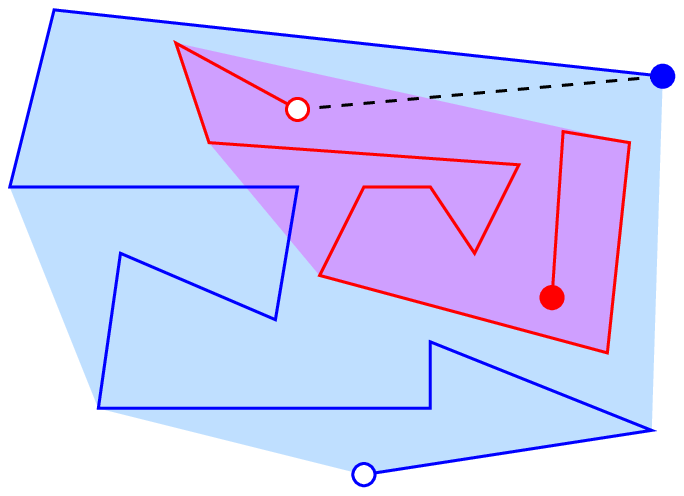,height=1.25in}\\(a)\end{tabular}
\hfil
\begin{tabular}{c}\epsfig{file=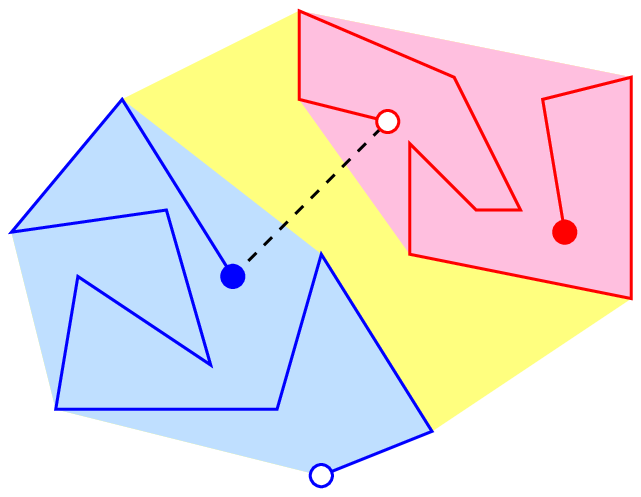,height=1.25in}\\(b)\end{tabular}
\hfil
\begin{tabular}{c}\epsfig{file=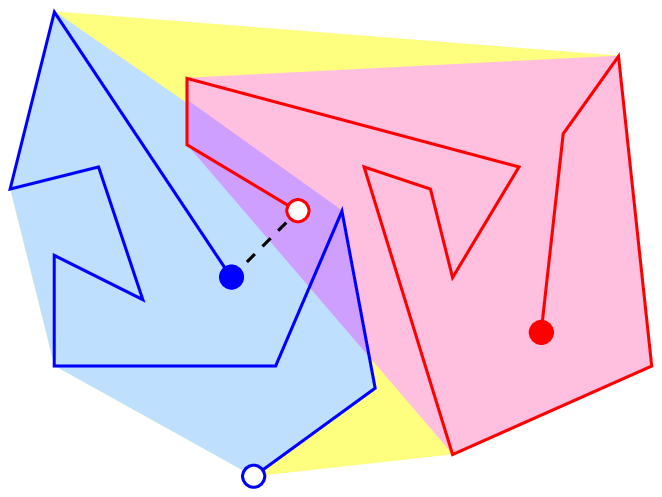,height=1.25in}\\(c)\end{tabular}
\caption{Convex hulls of adjacent subchains must be either (a)~nested,
(b)~disjoint, or (c)~overlapping with two common boundary points.
Hollow and solid circles mark respectively the first and last vertices
of each subchain.}
\label{Fig/subchains}
\end{figure}

If $\conv(A)$ and $\conv(B)$ are nested, then one of them is the
convex hull of $C$.  In general, deciding whether to given convex
polygons are nested requires $\Omega(n)$ time, but the special
structure of our problem allows a faster solution.  We define the
\emph{entrance} and \emph{exit} of a polygonal chain $C$ as follows.
The entrance of $C$ is a pair of rays whose common basepoint is the
first vertex of $C$ that is also a vertex of $\conv(C)$; the rays
contain the convex hull edges on either side of this vertex.  The exit
of $C$ is a similar pair of rays based at the last vertex of $C$ that
is also a vertex of $\conv(C)$.  See Figure~\ref{Fig/entrance}.

\begin{figure}[htb]
\centering\epsfig{file=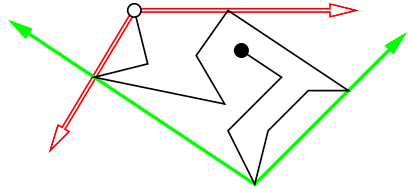,height=1in}
\caption{The entrance and exit of a polygonal chain.}
\label{Fig/entrance}
\end{figure}

Let $a$ be the last vertex of $A$, and let $b$ be the first vertex of
$B$.  The segment $ab$ does not intersect~$B$, so if $a$ is outside
the convex hull of~$B$, then $a$ must be outside the entrance of $B$
(\ie, on the opposite side of the entrance from~$B$).  More generally,
$\conv(A) \subset \conv(B)$ if and only if $\conv(A)$ lies completely
inside the entrance of $B$.\footnote{If $b$ is not a vertex of
$\conv(B)$, we can simplify the entrance of $B$ to a line through just
one convex hull edge.  While this modification simplifies our
algorithm somewhat, it does not significantly improve the running time.}
Similarly, $\conv(B) \subset \conv(A)$ if and only if $\conv(B)$ lies
completely inside the exit of $A$.  We can test in $O(\log n)$ time
whether a convex polygon (represented as an array of vertices in
counterclockwise order) lies inside a wedge.  Thus, if we can compute
the entrance and exit of any chain given those of its children, then
we can test for nested sibling subhulls in $O(\log n)$ time.
Fortunately, this is quite easy.  If both edges of $\conv(A)$ defining
the entrance of $B$ are also edges of $\conv(C)$, then the entrance of
$C$ is just the entrance of $A$.  Otherwise, the entrance of $C$
contains one of the two outer common tangents between $\conv(A)$ and
$\conv(B)$.

Now suppose $\conv(A)$ and $\conv(B)$ are not nested.  Using an
algorithm of Chazelle and Dobkin~\cite{cd-icott-87}, we can decide in
$O(\log n)$ time whether $\conv(A)$ and $\conv(B)$ intersect.  If the
two convex hulls have disjoint interiors, their algorithm also returns
a separating line~$\ell$.\footnote{Chazelle and Dobkin's algorithm
returns a pair of parallel separating lines, one tangent to each
polygon, but this is unnecessary for our result.  See
also~\cite{e-cedbt-85}.}  If we use $\ell$ as a local vertical
direction, we can divide $\conv(A)$ and $\conv(B)$ into separate upper
and lower hulls, such that one outer common tangent joins the two
upper hulls and the other joins to two lower hulls.  This is precisely
the setup required by the algorithm of Overmars and van Leeuwen, which
finds these two common tangents in $O(\log n)$ time \cite{ol-mcp-81}.

Finally, suppose the boundaries of $\conv(A)$ and $\conv(B)$ intersect
at two points.  In this case, we can find the two outer common tangent
lines between them, and thus compute $\conv(C)$, in $O(\log^2 n)$
time.  To find (say) the upper common tangent of $A$ and $B$, we
perform a modified binary search over the vertices of $\conv(A)$.  At
each step of this binary search, we find the upper tangent line $\ell$
to $\conv(B)$ (if any) passing through a vertex $a\in\conv(A)$ in
$O(\log n)$ time, using a second-level binary search.

Thus, we can compute the convex hull, entrance, and exit of $C$ from
the convex hulls, entrances, and exits of $A$ and $B$ in $O(\log^2 n)$
time.  By our earlier argument, it follows that we can maintain the
convex hull of $P$ in $O(\log^4 n)$ time per flipturn.  We can build
the original data structure in $O(n\log n)$ time by explicitly
computing the convex hulls of each subchain, each in linear time.
\end{proof}

\begin{theorem}
Given a simple $n$-gon $P$, a convexifying sequence of flipturns can
be computed in $O(L\log^4 n)$ time, where $L$ is the length of the
computed sequence.
\end{theorem}

\begin{proof}
We can construct the data structures to maintain the polygon and its
convex hull in $O(n\log n)$ time.  In addition to the convex hull
itself, we maintain a separate list of the lids of $P$, which requires
only trivial additions to our data structures.  This allows us to
choose a legal flipturn in constant time.  By Lemma \ref{convex}, we
can maintain both the polygon and its convex hull in $O(\log^4 n)$
time per flipturn.
\end{proof}

This theorem has an immediate corollary, using the results of Ahn
\etal~\cite{abchkm-fyl-00} and our Theorems~\ref{bracket} and
\ref{bracket2}.

\begin{corollary}
Given an $s$-oriented $n$-gon, a convexifying sequence of flipturns can
be computed in $O(sn\log^4 n)$ time.  In particular, for orthogonal
polygons, a convexifying flipturn sequence can be computed in
$O(n\log^4 n)$ time.
\end{corollary}

For orthogonal polygons, we can modify our algorithm to find a
flipturn sequence satisfying Theorem \ref{upper}, still in $O(n \log^4
n)$ time.  We maintain the diagonal pockets and orthogonal lids of $P$
in separate lists.  If there is a diagonal pocket, we flipturn it.
Otherwise, if some edge of the bounding box contains more than one
lid, we flipturn one of those pockets.  If each edge of the bounding
box has at most one lid, we can check whether any of these pockets is
bad in $O(\log^4 n)$ time.  To check one pocket, we flipturn it and
count diagonal pockets; if there is only one, we flipturn that and
count again.  If the pocket is bad and the original polygon has any
unchecked pockets, we undo the flipturn(s) and try the next pocket.
Each bad flipturn requires at most seven flipturns and six
anti-flipturns.  The proof of Theorem \ref{upper} ensures that we
perform at most $\floor{(n-4)/6}$ bad flipturns, so the total number
of data structure updates is at most $17(n-4)/6 = O(n)$.

It seems quite likely that our data structure can be improved.  One
obvious bottleneck in our algorithm is finding common tangents between
intersecting convex pseudo-disks, which currently takes $O(\log^2 n)$
time.  The more recent dynamic convex hull results of Chan
\cite{c-dpcho-99} and Brodal and Jakob \cite{bj-dpcho-00} may also be
useful here.  On the other hand, we are unable to prove even an
$\Omega(n\log n)$ lower bound, even for arbitrary polygons.  What is
the true complexity of computing flipturn sequences?


\section{Order Doesn't Matter}
\label{place}

Joss and Shannon showed that any simple polygon $P$ can be transformed
into a convex polygon by a sufficiently long sequence of flipturns.
If we always direct polygon edges so that they form a counterclockwise
cycle, then flipturns do not change the direction of any edge.  Since
flipturns also do not change edge lengths, the final convex shape of
$P$ is the same for any convexifying flipturn sequence.  We can easily
compute this shape in $O(n\log n)$ time by sorting the edges of $P$ by
their orientation.  For $s$-oriented polygons, this requires only
$O(n\log s)$ time.

In this section, we show that the \emph{position} of the final convex
polygon is also independent of the flipturn sequence.  To prove this
result, it suffices to show that we can predict the $y$-coordinate of
the top edge of the final convex polygon's bounding box.  The position
of the left edge follows from a symmetric argument, and these two
edges determine the polygon's final position.

We prove our theorem by induction on the number of
flipturns.\footnote{The results in this section actually hold for a
wider class of pivots called \emph{generalized flipturns}.  A
generalized flipturn rotates a chain of edges 180 degrees around the
midpoint of its endpoints without introducing self-intersections.
Generalized flipturns include standard, extended, and modified
flipturns as special cases.} Let $P$ be a non-convex polygon, let $ab$
be a lid of some pocket in $P$, and let $c$ be the midpoint of $ab$.
We subdivide the plane into horizontal strips using the horizontal
line $\ell_0$ through $c$, the horizontal lines $L$ passing through
every vertex of $P$, and the reflection $L'$ of $L$ across $\ell_0$.
Number the strips $1,2,3,\dots$ counting upwards from $\ell_0$ and
$-1,-2,-3,\dots$ counting downwards from $\ell_0$.  With this
numbering, any strip $i$ is the reflection of strip $-i$ across
$\ell_0$.  In particular, strips $i$ and $-i$ have the same width,
which we denote~$w_i$.  There are at most $2n+2$ strips altogether.

These strips subdivide the exterior of $P$ into trapezoidal regions.
We classify these trapezoids into several groups.  If a region is
unbounded, we call it an \emph{outer region}; otherwise, we call it an
\emph{inner region}.  We further classify outer regions into the
infinite \emph{strips} above or below $P$ (including the top and
bottom halfplanes), and the semi-infinite \emph{side regions} to the
left or right of $P$.  We also classify inner regions as
\emph{up-regions} and \emph{down-regions} as follows.  Consider the
shortest path through the exterior of $P$ starting at a point in the
interior of some inner region $\rho$ and ending at a point in some
outer region.  If the first segment of this path goes up from the
starting point, $\rho$ is an up-region; otherwise, $\rho$ is a
down-region.  We emphasize that this classification is independent of
the starting point within~$\rho$.  We show below that the total height
of the up regions is precisely the distance between the top of the
current polygon's bounding box and the top of the final convex
polygon's bounding box.

\begin{figure}[htb]
\centerline{\footnotesize\sf
    \begin{tabular}{c@{\qquad\qquad}c}
	\epsfig{file=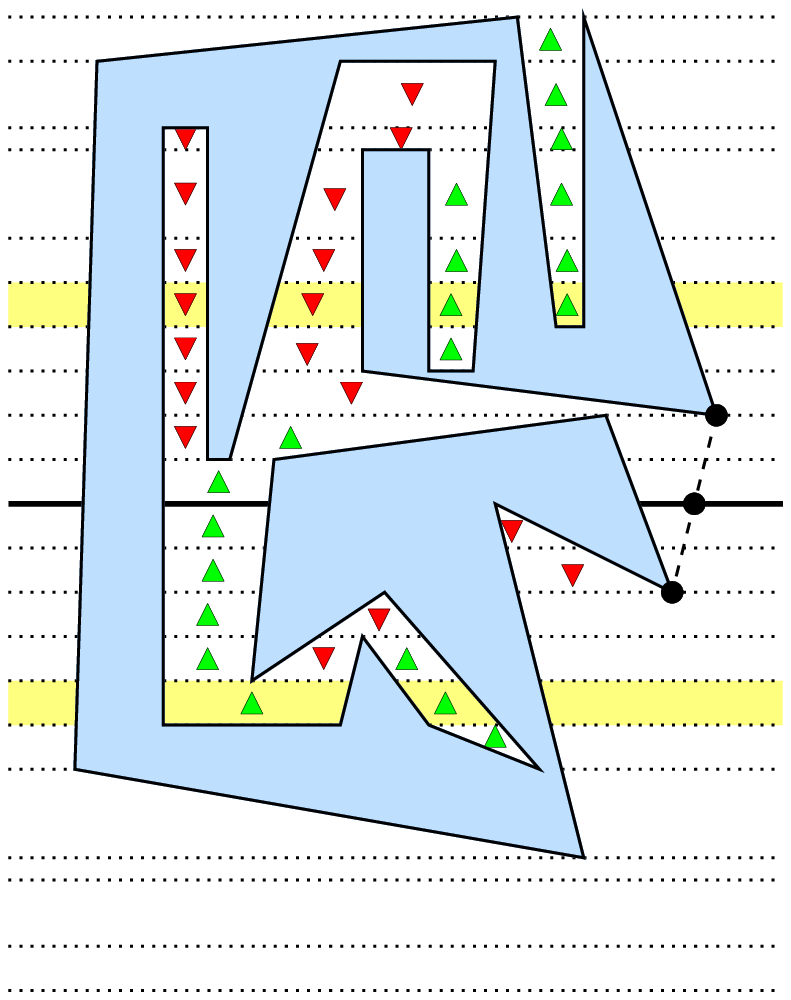,height=2.75in} &
	\epsfig{file=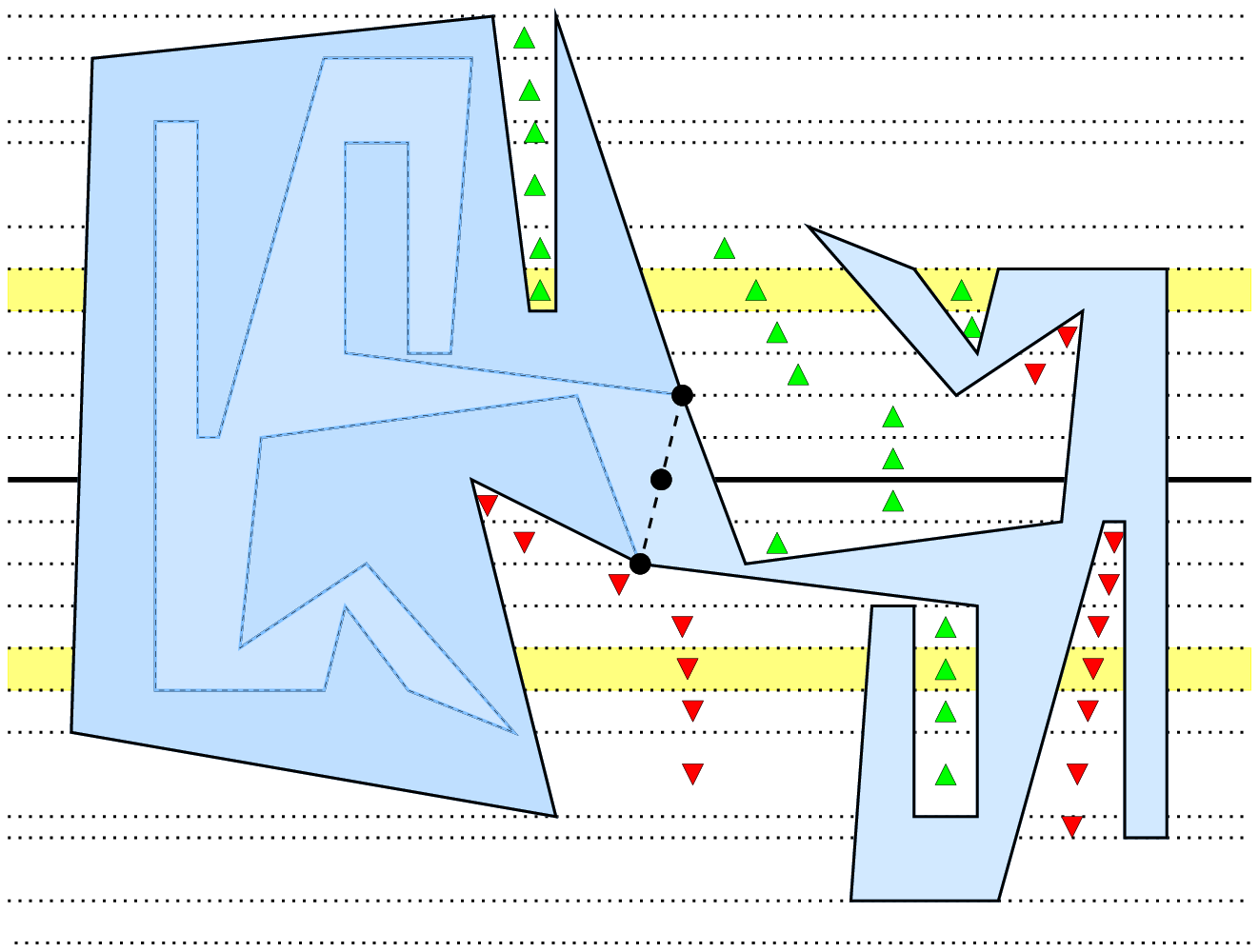,height=2.75in} \\
	(a) & (b)
    \end{tabular}}
\caption{Strips defined by a polygon and one of its pockets.  Strips
$5$ and $-5$ are highlighted.  Triangles indicate up-regions and
down-regions.  (a)~The original polygon $P$, with $u_5 = 4$ and $x_5 =
1$.  (b)~The flipturned polygon $P'$, with $u'_5 = 4$ and $x'_5 = 1$.}
\label{Fig/regions}
\end{figure}

For each $i>0$, let $u_i$ denote the number of up-regions in strips
$i$ and $-i$, and let $x_i$ be the indicator variable equal to $1$ if
strip $i$ intersects $P$ and $0$ otherwise.  See
Figure~\ref{Fig/regions}(a).

Let $P'$ be the result of flipturning the pocket $ab$.  This flipturn
moves any point on the boundary of the pocket from some strip $i$ to
the corresponding strip~$-i$.  The strips subdivide the exterior
of~$P'$ into regions exactly as the exterior of $P$, and we define the
corresponding variables $u'_j$ and~$x'_j$ mutatis mutandis.  See
Figure~\ref{Fig/regions}(b).

Our core lemma is the following.

\begin{lemma}
\label{invariant}
For all $i>0$, $u_i + x_i = u'_i + x'_i$.
\end{lemma}

\begin{proof}
Fix an index $i>0$.  We prove the theorem by induction on the number
of inner regions in the flipturned pocket.  If the pocket contains no
inner regions, it must be $y$-monotone, and we easily observe that
$u_i=u'_i$ and $x_i=x'_i$.

\def\tP{\tilde{P}}
\def\tu{\tilde{u}}
\def\tx{\tilde{x}}
\def\ts{\tilde{\sigma}}

The inner regions of $P$ have a natural forest structure, defined by
connecting each region to the next region encountered on a shortest
path to infinity.  The roots of this forest are inner regions directly
adjacent to outer regions, and its leaves are inner regions adjacent
to only one other region.  Let $\rho$ be some leaf region inside
pocket $ab$, let $\tP = P\cup\rho$, and define $\tu_i$ and $\tx_i$
analogously to $u_i$ and $x_i$ for this new polygon.  Finally, let
$\tP'$ be the result of flipturning the now-simpler pocket $ab$, let
$\rho'$ be the image of $\rho$ under this flipturn (so $\tP' =
P'\setminus \rho'$), and define $\tu'_i$ and $\tx'_i$ analogously.
Since $\tP$ has one less inner region than $P$, the inductive
hypothesis implies that $\tu_i + \tx_i = \tu'_i + \tx'_i$.

It suffices to consider the case where $\rho$ lies either in strip $i$
or in strip $-i$, since otherwise we have $\tu_i=u_i$, $\tx_i=x_i$,
$\tu'_i=u'_i$, and $\tx'_i=x'_i$, and so there is nothing to prove.
Moreover, if $\rho$ is in strip $i$, then $x_i = \tx_i = x'_i = \tx'_i
= 1$.

Suppose $\rho$ is an up-region, so $\tu_i = u_i-1$.  Some region
$\ts'$ of $\tP'$ is split into two regions by $\rho'$.  If we imagine
a continuous transformation from $\tP'$ to $P'$, the trapezoid $\rho'$
grows upward from the bottom edge of $\ts'$.  We have four cases to
consider, illustrated in the first two rows of
Figure~\ref{Fig/cases}.  (The last row shows the corresponding cases
when $\rho$ is a down-region.)

\begin{figure}[htb]
\centerline{\epsfig{file=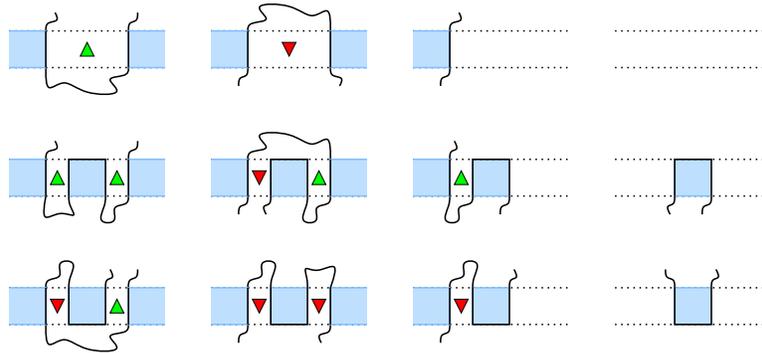,height=1.85in}}
\caption{Cases for the proof of Lemma \ref{invariant}.  From left to
right: $\ts'$ is an up-region, a down-region, a side region, or a
strip.  From top to bottom: $\ts'$ alone, split by the flipturned
up-region $\rho'$, or split by the flipturned down-region $\rho'$.}
\label{Fig/cases}
\end{figure}
\unskip
\begin{description}
\unskip
\item[\sl Case 1: $\ts'$ is an up-region.]
In this case $\rho'$ splits $\ts'$ into two up-regions, so $u'_i =
\tu'_i+1$.  If $\rho$ is in strip $-i$, then $\rho'$ is in strip $i$,
so $\tx'_i = x'_i = 1$ and $x_i = \tx_i$ (but these might be either
$0$ or $1$).

\item[\sl Case 2: $\ts'$ is a down-region.]
In this case $\rho'$ splits $\ts'$ into an up-region and a
down-region, so $u'_i = \tu'_i+1$.  If $\rho$ is in strip $-i$, then
$\tx'_i = x'_i = 1$ and $\tx_i = x_i$.

\item[\sl Case 3: $\ts'$ is a side region.]
In this case $\rho'$ splits $\ts'$ into an up-region and a side
region, so $u'_i = \tu'_i+1$.  If $\rho$ is in strip $-i$, then
$\tx'_i = x'_i = 1$ and $\tx_i = x_i$.

\item[\sl Case 4: $\ts'$ is a strip.]
Since $\rho$ is an up-region, $\tP'$ must touch the bottom edge
of~$\rho'$, which means that $\rho'$ must lie above $\tP$.  In this
case $\rho'$ splits $\ts'$ into two side regions, so $u'_i = \tu'_i$.
Since $\rho'$ is in strip $i$, we have $x_i = \tx_i = \tx'_i = 0$ but
$x'_i = 1$.
\end{description}

The lemma holds in every case.  Four similar cases arise when $\rho$
is a down-region and $\tu_i = u_i$.  In each case, we have $\tu'_i =
u'_i$, $\tx_i = x_i$, and $\tx'_i = x'_i$.  We omit further details.
\end{proof}

\begin{theorem}
\label{position}
The final convexified position of a polygon is independent of the
convexifying flipturn sequence and can be determined in $O(n)$ time.
\end{theorem}

\begin{proof}
Let $w_i$ denote the vertical width of strip $i$ (and strip $-i$).
Lemma \ref{invariant} implies that
\begin{equation}
\label{uxw}
	\sum_{i>0} (u_i + x_i)w_i = \sum_{i>0} (u'_i + x'_i)w_i.
\end{equation}
Let $\hat{y}$ and $\hat{y}'$ denote the $y$-coordinates of the top of
$P$ and $P'$, respectively, and let $y_0$ be the $y$-coordinate of the
lid midpoint $c$.  We easily observe that
\begin{equation}
\label{xw}
	\sum_{i>0} x_i w_i = \hat{y} - y_0
	\qquad\text{and}\qquad
	\sum_{i>0} x'_i w_i = \hat{y}' - y_0.
\end{equation}
Finally, define $U = \sum_{i>0} u_i w_i$ and $U = \sum_{i>0} u'_iw_i$.
Combining equations \eqref{uxw} and \eqref{xw}, we obtain the identity
$U + \hat{y} = U' + \hat{y}'$.  In other words, the total height of
all the up-regions plus the maximum $y$-coordinate of the polygon is
an invariant preserved by any flipturn.

Let $P^*$ be the convex polygon produced by some sequence of flipturns
starting from $P$, and define $U^*$ and $\hat{y}^*$ analogously to $U$
and $\hat{y}$.  Obviously, $P^*$ has no up-regions, so $U^* = 0$.
Thus, by induction on the number of flipturns, we have the identity
$\hat{y}^* = U + \hat{y}$.  Since $U+\hat{y}$ is independent of the
convexifying flipturn sequence, so is the vertical position of $P^*$.

We can compute $U$ in linear time by computing a horizontal
trapezoidal decomposition of $P$, using Chazelle's
algorithm~\cite{c-tsplt-91a} or its recent randomized variant by
Amato, Goodrich, and Ramos~\cite{agr-ltptm-00}, and then performing a
depth-first search of its dual graph.

The argument for the horizontal position of $P^*$ is symmetric.
\end{proof}


\section{The Worst Order Is Hard to Find}
\label{worst}

\begin{theorem}
Computing the longest standard or extended flipturn sequence for a
simple polygon is NP-hard.
\end{theorem}

\begin{proof}
It suffices to consider the special case of orthogonal polygons.  A
flipturn sequence for an orthogonal polygon has length greater than
$(n-4)/2$ if and only if it contains an orthogonal flipturn.  Thus, to
prove the theorem, we only need to show the NP-hardness of the
decision problem \textsc{Orthogonal Flipturn}: Given an orthogonal
polygon, does \emph{any} flipturn sequence contain an orthogonal
flipturn?  We prove this problem is NP-complete by a reduction from
\textsc{Subset Sum}: Given a set of positive integers $A = \set{a_1,
a_2, \dots, a_n}$ and another integer $T$, does any subset of $A$ sum
to $T$?  The reduction algorithm is given in Figure~\ref{Fig/algo}.
The algorithm constructs a polygon in linear time by walking along its
edges in clockwise order, starting and ending at the top of the first
step.  (The algorithm assumes without loss of generality that $n$ is
even.)  Figure~\ref{Fig/nphard} shows an example of the reduction.

\begin{figure}[htb]
\centering\small
\begin{algo}
\underline{$\mathsc{SubsetSum}(A, T) \mapsto \mathsc{OrthogonalFlipturn}$:}\+\\
	\Comment{Upper steps and inward spikes}\\
	for $i \gets 1$ to $n/2$	\+\\
	    \textsc{South}$(a_{2i-1})$; 
	    \textsc{East}$(a_{2i-1})$;		
	    \textsc{South}$(1)$;	\\
	    \textsc{West}$(T+2n-4i+4)$;	
	    \textsc{South}$(1)$;	
	    \textsc{East}$(T+2n+4i-4)$	\-
\\[1ex]
	\Comment{Test spike}\\
	\textsc{South}$(T+2)$;		
	\textsc{East}$(1)$;		
	\textsc{North}$(T)$;		
	\textsc{East}$(1)$;		
	\textsc{South}$(T+1)$;
	\textsc{West}$(2)$;
\\[1ex]
	\Comment{Lower steps and outward spikes}\\
	for $i \gets 1$ to $n/2$	\+\\
	    \textsc{South}$(1)$;		
	    \textsc{East}$(a_{n-2i+2})$;	
	    \textsc{South}$(a_{n-2i+2})$\\	
	    \textsc{East}$(T+4i+2)$;	
	    \textsc{South}$(1)$;	
	    \textsc{West}$(T+4i+2)$;	\-
\\[1ex]
	\Comment{Close off the polygon}\\
	$\Sigma \gets \sum_{i=1}^n a_i$		\\
	\textsc{West}$(T + \Sigma + 2n + 2)$;
	\textsc{North}$(T + \Sigma + 2n + 3)$;
	\textsc{East}$(T + 2n + 2)$
\end{algo}
\unskip
\caption{The algorithm to reduce \textsc{SubsetSum} to
\textsc{Orthogonal Flipturn}.}
\label{Fig/algo}
\end{figure}

\begin{figure}[htb]
\centerline{\footnotesize\sf
\psfrag{a1}{$a_1$}\psfrag{a2}{$a_2$}\psfrag{a3}{$a_3$}\psfrag{a4}{$a_4$}
\psfrag{T}{$T$}\psfrag{a1+a2+a4=T}[br][br]{$a_1+a_2+a_4=T$}
\begin{tabular}{c}
\small	\epsfig{file=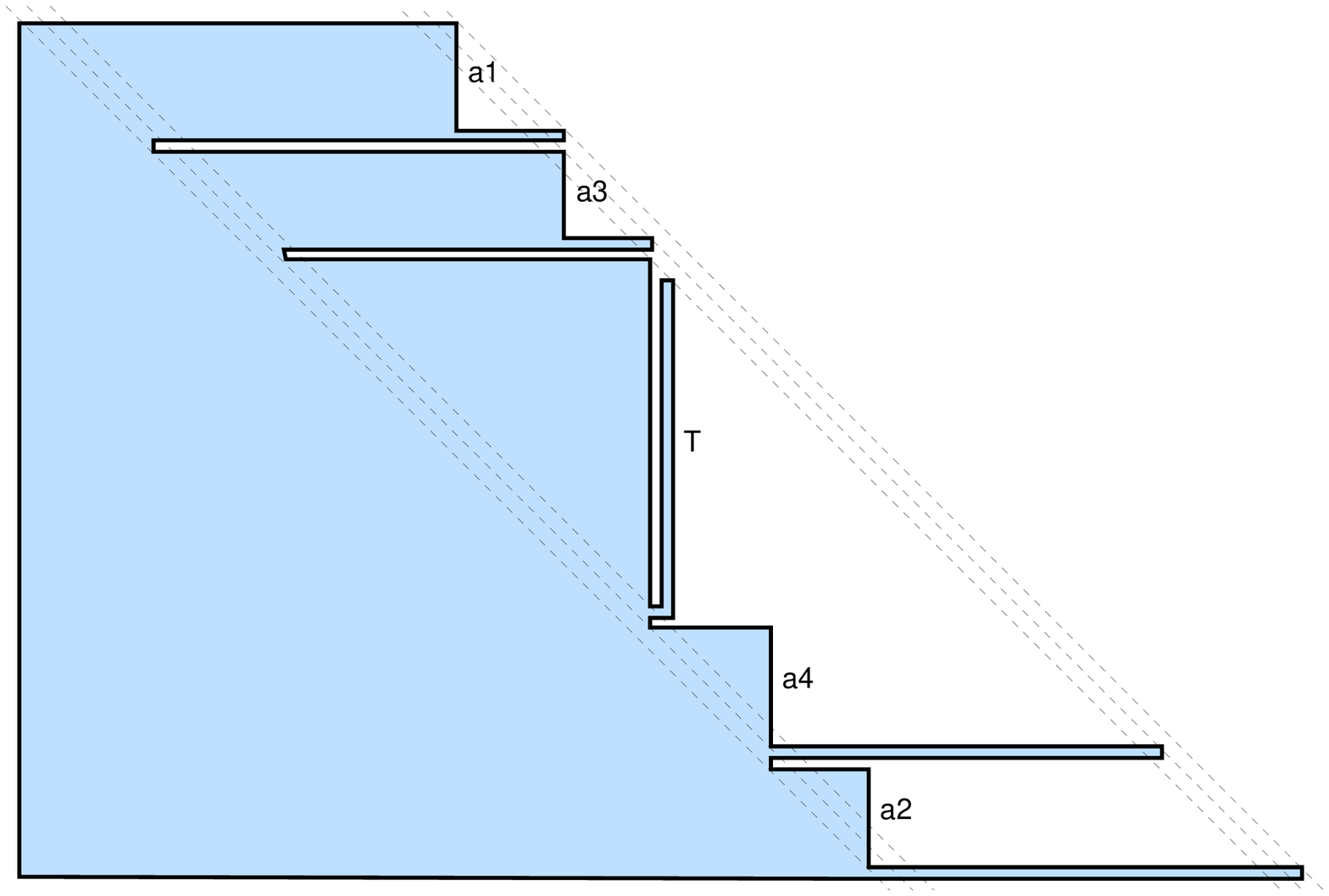,height=3in}\\(a)
	\\[3ex]
	\epsfig{file=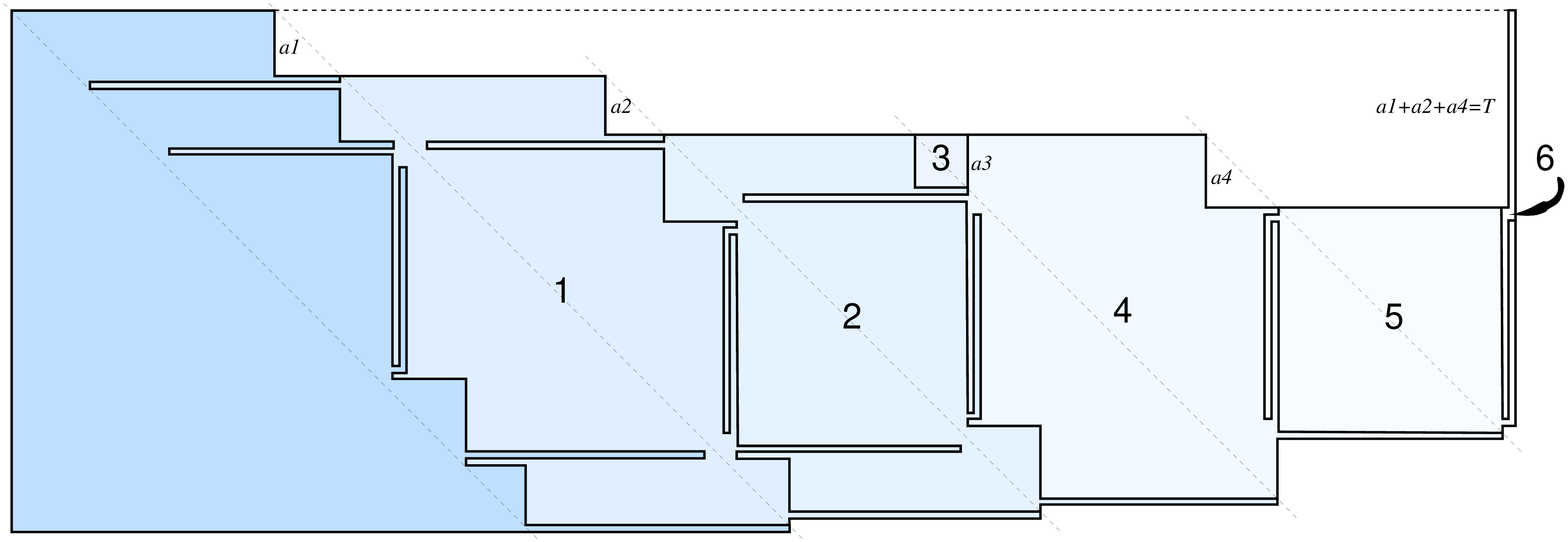,height=2in}\\(b)
\end{tabular}}
\caption{The reduction from \textsc{Subset Sum} to \textsc{Orthogonal
Flipturn}.  (a) Storing the set $\set{a_1, a_2, a_3, a_4}$ and the
target sum $T$.  (b) If we flipturn the step of height $a_3$ as soon
as possible (flipturn 3) and leave the other steps alone, then
flipturning the test spike (flipturn 6) creates an orthogonal pocket,
since $a_1 + a_2 + a_4 = T$.}
\label{Fig/nphard}
\end{figure}

The basic structure of the polygon is a staircase, with one square
step for each of the $a_i$, plus one long step of height $T$ splitting
the other steps in half.  Just below each of the upper steps is an an
inward horizontal spike; just above each of the lower steps is an
outward horizontal spike; and just behind the long step is a vertical
\emph{test spike} of length exactly~$T$.  The horizontal spikes all
have length greater than $T$, and they increase in length as they get
closer to the top and bottom of the polygon.  

At any point during the flipturning process, the polygon has one main
pocket containing the test spike and several secondary pockets
containing one or more smaller steps, each of whose heights is some
$a_i$.  Initially, there is just one secondary pocket, of height and
width~$a_1$.  The $i$th step (\ie, the one with height $a_i$) is
exposed the $(i-1)$th time the main pocket is flipturned.  No matter
which flipturns we perform before flipturning the test spike, the
vertical distance $\Delta$ between the top endpoint of the main
pocket's lid and the top edge of the polygon's bounding box is always
the sum of elements of~$A$.  Specifically, if we flipturn every step
whose size is an element of some subset $B\subseteq A$ as soon as it
becomes available, then just before the test spike is flipped,
$\Delta$ is the sum of the elements of $A\setminus B$; see
Figure~\ref{Fig/nphard}(b).  Thus, since the test spike has length
$T$, flipturning it can create an orthogonal pocket if and only if
some subset of $A$ sums to~$T$.
\end{proof}

Note that the polygon produced by our reduction never has more than
one orthogonal pocket; the longest flipturn sequence has either
$(n-4)/2$ or $(n-2)/2$ flipturns.  Thus, even approximating the
maximum number of orthogonal flipturns is NP-hard.

Our reduction only proves that finding the longest flipturn sequence
is \emph{weakly} NP-hard.  In particular, it says nothing about
\emph{lattice} polygons in their standard representation as a cycle of
unit-length orthogonal segments.  We conjecture that for such
polygons, there is a polynomial time dynamic programming algorithm,
similar to the $O(nT)$ algorithm for \textsc{SubsetSum}.

Finally, how hard is it to find the \emph{shortest} sequence of
flipturns that convexifies a given simple polygon?  It seems unlikely
that the question ``Does \emph{every} flipturn sequence have an
orthogonal flipturn?'' is NP-hard.

\subsection*{Acknowledgments}

All but one of the authors thank Godfried Toussaint for organizing
the Barbados workshop where most of this work was done.  Thanks also
to David Bremner, Ferran Hurtado, Vera \Sacristan, and Mike Soss for
sharing coffee, rum, and ideas.

\clearpage
\bibliographystyle{abuser}
\bibliography{flipturn,geom}

\end{document}